\documentclass[aps,prb,twocolumn,superscriptaddress,showpacs]{revtex4-1}
\usepackage{amsmath}
\usepackage{color}
\usepackage{dcolumn}
\usepackage{amssymb}
\usepackage{epsfig}
\usepackage{multirow}
\usepackage{amsbsy}
\usepackage{array}
\usepackage{hyperref}
\usepackage{bm} 
\usepackage{extarrows}
\usepackage{graphicx}
\usepackage{appendix}

\begin{document}

\title{Hidden $SU(2)$ Symmetries, the Symmetry Hierarchy and the Emergent Eight-Fold Way in Spin-1 Quantum Magnets}
 
\author{Hui-Ke Jin}
\affiliation{Department of Physics, Zhejiang University, Hangzhou 310027, China}

\author{Jian-Jian Miao}
\affiliation{Kavli Institute for Theoretical Sciences, University of Chinese Academy of Sciences, Beijing 100190, China}

\author{Yi Zhou}
\affiliation{Department of Physics, Zhejiang University, Hangzhou 310027, China}
\affiliation {Beijing National Laboratory for Condensed Matter Physics $\&$ Institute of Physics, Chinese Academy of Sciences, Beijing 100190, China}
\affiliation{CAS Center for Excellence in Topological Quantum Computation, University of Chinese Academy of Sciences, Beijing 100190, China}
\affiliation{Collaborative Innovation Center of Advanced Microstructures, Nanjing University, Nanjing 210093, China}

\date{\today}

\begin{abstract}
The largest allowed symmetry in a spin-1 quantum system is an $SU(3)$ symmetry rather than the $SO(3)$ spin rotation. In this work, we reveal some $SU(2)$ symmetries as subgroups of $SU(3)$ that, to the best of our knowledge, have not previously been recognized. Then, we construct $SU(2)$ symmetric Hamiltonians and explore the ground-state phase diagram in accordance with the $SU(3)\supset SU(2)\times U(1)$ symmetry hierarchy. 
It is natural to treat the eight generators of the $SU(3)$ symmetry on an equal footing; this approach is called the eight-fold way.
We find that the spin spectral functions and spin quadrupole spectral functions share the same structure, provided that the elementary excitations are flavor waves at low energies, which serves as a clue to the eight-fold way. An emergent $S=1/2$ quantum spin liquid is proposed to coexist with gapful spin nematic order in one of the ground states. In analogy to quantum chromodynamics, we find the gap relation for hydrodynamic modes in quantum spin-orbital liquid states, which is nothing but the Gell-Mann-Okubo formula.

\end{abstract}

\maketitle

\section{Introduction}
\label{sec:intro}
The symmetry principle plays a fundamental role with respect to the laws of nature. It provides an infrastructure and coherence for summarizing physical laws that are independent of any specific dynamics.
Noether's theorem says that every continuous symmetry of the action of a physical system is associated with a corresponding conservation law.
The standard paradigm for describing phase transitions and critical phenomena is Landau's theory of symmetry breaking. The states of matter are classified on the basis of symmetries. 
A higher-temperature phase is of a high symmetry characterized by a group $G$, while a lower-temperature phase is of a low symmetry characterized by a subgroup $H\subset G$. 
A low-energy effective theory can be constructed in terms of order parameters and is described by all terms that are allowed according to the relevant symmetries.
A hierarchy of symmetries is also widely used in particle physics to understand the dynamics of elementary particles.

Meanwhile, spin-1 quantum magnets are of great interest in physics. One famous example is the Haldane phase in one-dimensional (1D) spin-1 chains\cite{hgap},
in which fractional spin-$1/2$ end states are protected by the spin rotational symmetry in a phenomenon called symmetry-protected topological order\cite{spt1,eespt,spt2,spt3}.
Spin-1 systems are also able to host spin nematic orders in dimensions of $D>1$; such orders are characterized by long-range spin quadrupolar correlations, and the possibility for fractional spinon excitations to coexist with spin nematic orders has also been proposed\cite{s1fermi1}.
Such quantum magnets are widely encountered in various materials, especially transition metal compounds, in which a local $S=1$ magnetic moment can be formed in a cation via Hund's coupling; examples include $3d^{8}$ Ni$^{2+}$ and $3d^{6}$ Fe$^{2+}$.
In this work, we shall reveal several hidden $SU(2)$ symmetries in spin-1 quantum magnets in addition to spin rotational symmetry, and we will study spin-1 quantum systems with the help of the symmetry hierarchy. 

For a spin-1 quantum magnet, there are three local states, namely, $|S^{z}=\pm1\rangle$ and $|S^{z}=0\rangle$, and eight independent local Hermitian operators: three spin vector operators, $S^{x}$, $S^{y}$ and $S^{z}$, and five spin quadrupolar operators, 
\begin{eqnarray}
Q^{x^{2}-y^{2}} & = & (S^{x})^{2}-(S^{y})^{2},\nonumber \\
Q^{3z^{2}-r^{2}} & = & \frac{1}{\sqrt{3}}\left[2(S^{z})^{2}-(S^{x})^{2}-(S^{y})^{2}\right], \nonumber \\ 
Q^{xy} & = & S^{x}S^{y}+S^{y}S^{x},  \\
Q^{yz} & = & S^{y}S^{z}+S^{z}S^{y}, \nonumber \\
Q^{zx} & = & S^{z}S^{x}+S^{x}S^{z}. \nonumber
\end{eqnarray}
To illustrate the symmetry hierarchy, we consider a generic two-body interacting Hamiltonian as follows:
\begin{equation}\label{eq:Hgeneric}
\mathcal{H}=\sum_{\langle i,j\rangle}\Big(\sum_{\alpha\beta}J_{\alpha\beta}S^{\alpha}_{i}S^{\beta}_{j}+ \sum_{\mu\nu}J_{\mu\nu}Q^{\mu}_{i}Q^{\nu}_{j}+ \sum_{\alpha\mu}I_{\alpha\mu}S^{\alpha}_{i}Q^{\mu}_{j}\Big),
\end{equation}
where $\langle i,j\rangle$ is a pair of nearest neighboring sites; $\alpha$ and $\beta$ denote $x$, $y$, and $z$; and $\mu$ and $\nu$ denote $x^{2}-y^{2}$, $3z^{2}-r^{2}$, $xy$, $yz$, and $zx$.
The $SO(3)$ spin rotational symmetry is achieved when $I_{\alpha\mu}=0$, $J_{\alpha\beta}=\delta_{\alpha\beta}J_{1}$ and $J_{\mu\nu}=\delta_{\mu\nu}J_{2}$. Furthermore, $\mathcal{H}$ will be $SU(3)$ symmetric when $J_1=J_2$, and the $SU(3)$ group is generated by eight operators $\{\bm{S},\bm{Q}\}$\cite{bbq1d4,SU3ULSL,SU3ULSS}.
The $SO(3)$ model is well studied: a phase diagram consisting of a ferromagnetic phase, a dimerized phase, Haldane phases and a critical phase has been constructed in one dimension\cite{bbq1d1,bbq1d2,bbq1d3,gutzsu3}, and the $SO(3)$ model can host spin nematic ground states in dimensions of $D>1$\cite{s1fermi1,bbq2d1,bbq2d2,bbq2d3,bbq2d4,bbq2d5,bbq2d6,bbq2d7,bbq2d8}.

The model defined in Eq.~\eqref{eq:Hgeneric} has typically been studied in accordance with the $SU(3)\supset SO(3) \cdots$ symmetry hierarchy. 
Nevertheless, there are other $SU(2)$ subgroups belonging to the $SU(3)$ group, 
and this fact implies the existence of a slice of $SU(2)$ symmetries in addition to the $SO(3)$ spin rotation in spin-1 quantum magnets of which, to the best of our knowledge, the research community is not aware. This situation inspires us to search for Hamiltonians that respect these hidden symmetries;
for this purpose, a new symmetry hierarchy, $SU(3)\supset SU(2)\times U(1)\cdots$, will be adopted to reveal novel states with various low-energy excitations. To describe these states, it is natural to treat all operators $\{\bm{S},\bm{Q}\}$ on an equal footing, which is reminiscent of the ``eight-fold way" in quantum chromodynamics (QCD).

The paper is organized as follows. We reveal hidden $SU(2)$ symmetries  and construct corresponding $SU(2)$ symmetric Hamiltonians in Section~\ref{sec:sym.H}. In the spirit of eight-fold way, we apply flavor-wave mean-field theory to study the $SU(2)\times U(1)$ model and demonstrate that the similar structure in spin spectral functions and spin quadrupole spectral functions serves a clue to the eight-fold way in Section~\ref{sec:MFT}.
In Section~\ref{sec:beyondMFT}, we go beyond the mean-field theory and find an emergent $S=1/2$ gapless quantum spin liquid state coexisting with spin nematic order. We also find Gell-Mann-Okubo formula like gap relations for the hydrodynamic modes in quantum spin-orbial liquid states. 
Section~\ref{sec:summary} is devoted to summary.

\section{Hidden $SU(2)$ Symmetries and Hamiltonians}\label{sec:sym.H}

\subsection{Hidden $SU(2)$ Symmetries}
\label{sec:hsu2}
It turns out that there are three hidden $SU(2)$ symmetries in addition to well known spin rotational $SO(3)$ symmetry in a spin-1 system, which are generated as follows (see Appendix.\ref{app:SU2}): 
\begin{eqnarray}
SU(2)_{\alpha}&:&\{Q^{zx},S^{y},\frac{1}{2}Q^{x^2-y^2}+\frac{\sqrt{3}}{2}Q^{3z^2-r^2}\},\nonumber\\
SU(2)_{\beta}&:&\{Q^{yz},S^{x},\frac{1}{2}Q^{x^2-y^2}-\frac{\sqrt{3}}{2}Q^{3z^2-r^2}\},\nonumber\\
SU(2)_{\gamma}&:&\{Q^{xy},S^{z},Q^{x^2-y^2}\}.\label{eq:SU2SQ}
\end{eqnarray}
Each set of these generators consists of one component of the spin vector $\bm{S}$ and two components of the spin quadrupole $\bm{Q}$. 
Note that these three sets of generators are related to each other by the following cycle: $S^{x}\to S^{y}\to S^{z}\to S^{x}$. 
In the remaining part of this work, we shall focus on $SU(2)_{\gamma}$; $SU(2)_{\alpha}$ and $SU(2)_{\beta}$ can then be obtained in accordance with this cycle.

For the $SU(2)_{\gamma}$ symmetry, $S^{z}$ generates spin rotations along the $z$-axis, and the other two generators, $Q^{xy}$ and $Q^{x^2-y^2}$, correspond to {\em two-magnon} processes,
as can be seen from
\begin{subequations}
\begin{eqnarray}
Q^{x^2-y^2}&=&\frac{1}{2}\left[(S^{+})^2+(S^{-})^2 \right],\\
Q^{xy}&=&\frac{1}{2i}\left[(S^{+})^2-(S^{-})^2 \right],
\end{eqnarray}
\end{subequations}
where $S^{\pm}=S^{x}\pm iS^{y}$. Let us define,
\begin{subequations}
\begin{eqnarray}
J^z &=&\frac{1}{2}S^z, \\
J^+ &=&\frac{1}{2}(Q^{x^2-y^2}+iQ^{xy})=\frac{1}{2}(S^+)^2,\\
J^- &=&\frac{1}{2}(Q^{x^2-y^2}-iQ^{xy})=\frac{1}{2}(S^-)^2.
\end{eqnarray}
\end{subequations}
It is easy to verify that $\{J^z,J^{\pm}\}$ satisfy the $SU(2)$ Lie algebra.
Therefore, the spontaneous breaking of the $SU(2)_{\gamma}$ symmetry along the $S^{z}$ direction will give rise to {\em two-magnon} low-energy excitations,
while spontaneous symmetry breaking along the $Q^{xy}$ and $Q^{x^2-y^2}$ directions will give rise to an admixture of {\em one- and two-magnon} excitations, which will tend to restore the $SU(2)_{\gamma}$ symmetry.

The underlying $SU(3)$ structure and the hidden $SU(2)$ symmetries will be more transparent in the Cartesian representation of the spin states: $|x\rangle=i (|1\rangle-|-1\rangle)/\sqrt{2}$, $|y\rangle=(|1\rangle+|-1\rangle)/\sqrt{2}$, and $|z\rangle=-i|0\rangle$. 
Then, a spin state can be written as $|\bm{d}\rangle=d^{x}|x\rangle+d^{y}|y\rangle+d^{z}|z\rangle$, where $\bm{d}=\left(d^{x},d^{y},d^{z}\right)$ is a complex vector and the normalization condition is given by $|\bm{d}|^{2}=1$.
The expectation values for $\{\bm{S},\bm{Q}\}$ can be expressed in terms of $\bm{d}$ as follows:  
\begin{eqnarray}
\langle S^{\alpha}\rangle & = & -i\epsilon_{\alpha\beta\gamma}\bar{d}^{\beta}d^{\gamma}, \nonumber \\
\langle Q^{\alpha\beta} \rangle|_{\alpha\neq\beta} & = & -(\bar{d}^{\alpha}d^{\beta}+\bar{d}^{\beta}d^{\alpha}), \nonumber \\
\langle Q^{x^{2}-y^{2}}\rangle & = &|d^{y}|^{2}-|d^{x}|^{2}, \nonumber \\ 
\langle Q^{3z^{2}-r^{2}}\rangle & = & \frac{1}{\sqrt{3}}(2|d^{z}|^{2}-|d^{y}|^{2}-|d^{x}|^{2}),
\end{eqnarray}
where $\bar{d}^{\alpha}$ is the complex conjugate of $d^{\alpha}$ and $\epsilon^{\alpha\beta\gamma}$ is a three-rank antisymmetric tensor. Thus, a spin-1 quantum system can be described by the following path integral:
\begin{equation}\label{eq:pathint}
\mathcal{Z}=\int\mathcal{D}[\bm{d},\bm{\bar{d}}\,] \delta(|\bm{d}|^2-1) e^{-\int_0^{\beta}d\tau \left\{ \sum_{i}\bm{\bar{d}}_i\cdot \partial_{\tau}\bm{d}_i-\mathcal{H} \right\}},
\end{equation}
where the Hamiltonian $\mathcal{H}$ is given by Eq.~\eqref{eq:Hgeneric} with $\{\bm{S},\bm{Q}\}$ replaced with their expectation values.
Now, it is clear that all of the special unitary transformations of $\bm{d}$ give rise to the $SU(3)$ group and that the special unitary transformations of any two components of $\bm{d}$ lead to either $SU(2)_{\alpha}$, $SU(2)_{\beta}$ or $SU(2)_{\gamma}$.

\subsection{$SU(2)$-symmetric Hamiltonians}
\label{sec:su2ham}
Now, we are in a position to construct Hamiltonians in accordance with the $SU(2)_{\gamma}$ symmetry.
A generic spin-1 Hamiltonian can be written in terms of $\{\bm{S},\bm{Q}\}$ in a bilinear form as shown in Eq.~\eqref{eq:Hgeneric}.
Using group theory, one is able to obtain all $SU(2)_{\gamma}$-symmetric two-body interactions (see Appendix.\ref{app:SU2state}).
These $SU(2)_{\gamma}$-symmetric Hamiltonians are linear combinations of the following six terms:
\begin{subequations}\label{eq:H1-6}
\begin{eqnarray}
\mathcal{H}_{1} &=& \sum_{\langle i,j\rangle}S^{z}_{i}S^{z}_{j}+Q^{xy}_{i}Q^{xy}_{j}+Q^{x^{2}-y^{2}}_{i}Q^{x ^{2}-y^{2}}_{j},\\
\mathcal{H}_{2} &=& \sum_{\langle i,j\rangle}S^{x}_{i}S^{x}_{j}+Q^{yz}_{i}Q^{yz}_{j}+S^{y}_{i}S^{y}_{j}+Q^{zx}_{i}Q^{zx}_{j},\\
\mathcal{H}_{3} &=& \sum_{\langle i,j\rangle}Q^{3z^{2}-r^{2}}_{i}Q^{3z^{2}-r^{2}}_{j},\\
\mathcal{H}_{4} &=& \sum_{\langle i,j\rangle}D_{ij}\left[S^{x}_{i}S^{y}_{j}+Q^{yz}_{i}Q^{zx}_{j}-(i\leftrightarrow{}j)\right],\\
\mathcal{H}_{5} &=& \sum_{\langle i,j\rangle}D_{ij}\left[S^{y}_{i}Q^{zx}_{j}+Q^{yz}_{i}S^{x}_{j}-(i\leftrightarrow{}j)\right],\\
\mathcal{H}_{6} &=& \sum_{\langle i,j\rangle}D_{ij}\left[Q^{zx}_{i}S^{x}_{j}+Q^{yz}_{i}S^{y}_{j}-(i\leftrightarrow{}j)\right],
\end{eqnarray}
\end{subequations}
where the $D_{ij}=-D_{ji}=\pm 1$ define a direction along each bond $\langle i,j\rangle$ and are translationally invariant.
The Hamiltonians $\mathcal{H}_{1-6}$ can be classified with respect to the time-reversal symmetry $\mathcal{T}$, the spatial inversion symmetry $\mathcal{I}$, and additional symmetries as summarized in Table \ref{tab:H1-6}. 

Further discussions are presented as follows:
(1) $\mathcal{H}_{1}$ consists of the $SU(2)_{\gamma}$ generators $\{S^{z}_i,Q^{xy}_i,Q^{x^{2}-y^{2}}_i\}$ and can be viewed as a spin-1/2 Heisenberg model in the subspace spanned by the local basis $\{|x\rangle_i,|y\rangle_i\}$.  
(2) Let us define $\Lambda_8=\sum_{i}Q^{3z^{2}-r^{2}}_{i}$; then, we have $[\Lambda_8,\mathcal{H}_{1}]=[\Lambda_8,\mathcal{H}_{2}]=[\Lambda_8,\mathcal{H}_{3}]=[\Lambda_8,\mathcal{H}_{5}]=0$.  
Thus, $\mathcal{H}_{1-3}$ and $\mathcal{H}_{5}$ have an $SU(2)_{\gamma}\times U(1)$ symmetry, where the additional global $U(1)$ symmetry is generated by $\Lambda_8$.
(3) $\mathcal{H}_{1}$ has an additional local $U(1)$ symmetry generated by $Q_{i}^{3z^2-r^2}$.
(4) $\mathcal{H}_{3}$ has an additional local $U(2)$ symmetry that is generated by $\{S^{z}_i,Q^{xy}_i,Q^{x^{2}-y^{2}}_i\}$ and $Q^{3z^{2}-r^{2}}_{i}$.

\begin{table}[htbp]
\caption{Typical $SU(2)_{\gamma}$-symmetric Hamiltonians in Eq.~\eqref{eq:H1-6}, which are classified with respect to time reversal ($\mathcal{T}$), spatial inversion ($\mathcal{I}$), and additional global/local symmetries.}
\label{tab:H1-6}
\renewcommand\arraystretch{1.2}
\begin{tabular}{c|c|c|c|c}
\hline
Hamiltonian &  \hspace{0.5cm}$\mathcal{T}$\hspace{0.5cm} & \hspace{0.5cm}$\mathcal{I}$\hspace{0.5cm} & \hspace{0.6cm}Global\hspace{0.6cm} & \hspace{0.4cm}Local\hspace{0.4cm} \\
\hline
$\mathcal{H}_{1}$ & Yes & Yes & $SU(2)\times U(1)$ & $U(1)$\\
$\mathcal{H}_{2}$ & Yes & Yes & $SU(2)\times U(1)$ & \\
$\mathcal{H}_{3}$ & Yes & Yes & $SU(2)\times U(1)$ & $U(2)$\\
$\mathcal{H}_{4}$ & Yes & No & $SU(2)$ & \\
$\mathcal{H}_{5}$ & No & No & $SU(2)\times U(1)$ & \\
$\mathcal{H}_{6}$ & No & No & $SU(2)$ & \\
\hline
\end{tabular}
\renewcommand\arraystretch{1}
\end{table}

\section{Eight-fold way and flavor-wave mean-field theory: $SU(2)\times U(1)\times\cal{T}\times\cal{I}$ model}\label{sec:MFT}
Regarding the $SU(3)\supset SU(2)\times U(1)$ symmetry hierarchy, we would like to treat all of the $SU(3)$ generators $\{\bm{S},\bm{Q}\}$ on an equal footing.
In this spirit, we study ground states and low-energy excitations with the help of flavor-wave mean-field theory\cite{bbqmodel2,flavor1, flavor2,flavor3,flavor4}, and then discuss how the spectral functions can be exploited to detect the eight-fold way.

First, we minimize the energy functional $\mathcal{H}(\bm{d},\bm{\bar{d}})$ with respect to the complex fields $\bm{d}_i$ to determine the ground state.
Second, we assign three flavors of Schwinger bosons, $a_{n\alpha}(j)$, to each site $j$ on the $n^{th}$ sublattice, 
where $\alpha=x,y,z$ refers to the local spin states. For example, $n=1$ for a uniform state, while $n=1,2$ for a bipartite-lattice ordered state.
The operators $\{\bm{S},\bm{Q}\}$ can be written bilinearly in terms of the Schwinger bosons, and the physical Hilbert space can be restored by imposing a single-occupancy condition (see Appendix.\ref{app:FW}).
Third, without loss of generality, we let the Schwinger bosons condense at $a_{n\tilde{x}}$ to obtain ordered states, where $a_{n\tilde{x}}$ and the other two orthogonal components, $a_{n\tilde{y}}$ and $a_{n\tilde{z}}$, are related to $(a_{nx}, a_{ny}, a_{nz})$ by an $SU(3)$ rotation $\Omega_n$ as follows: $(a_{n\tilde{x}}, a_{n\tilde{y}}, a_{n\tilde{z}})^T = \Omega_{n}(a_{nx}, a_{ny}, a_{nz})^T$.
Such an $\Omega_{n}$ is determined by the mean-field vector $\bm{d}$ and enables us to attribute the condensate to $a_{n\tilde{x}}$ alone, while treating $a_{n\tilde{y}}$ and $a_{n\tilde{z}}$ as small fractions. Then, the low-energy Hamiltonian can be bilinearized by the Holstein-Primakoff transformation: $a_{n\tilde{x}}^{\dagger}(j) = a_{n\tilde{x}}(j) = \sqrt{M-a_{n\tilde{y}}^{\dagger}(j)a_{n\tilde{y}}(j)-a_{n\tilde{z}}^{\dagger}(j)a_{n\tilde{z}}(j)}$, 
where we will ultimately take $M=1$ for the single-occupancy case.
Expansion in $1/M$ and Bogoliubov transformation will give rise to a diagonalized Hamiltonian in $k$-space (see Appendix.\ref{app:FW}):
$\mathcal{H}=\sum_{m,\bm{k}}\omega_{m}(\bm{k})b^{\dagger}_{m}(\bm{k})b_{m}(\bm{k})+\mathcal{C}$,
where $\omega_{m}(\bm{k})$ is the energy dispersion of the $m$-th flavor-wave branch, $b_{m}(\bm{k})$ is a bosonic Bogoliubov quasiparticle, and $\mathcal{C}$ is a constant.
For a uniform state, $m=1,2$, while for a bipartite-lattice ordered state, $m=1,2,3,4$. 
As long as the vector $\bm{d}$ is given by the mean-field theory, we will be able to obtain $\omega_{m}(\bm{k})$ and $b_{m}(\bm{k})$ simultaneously.

In particular, we are interested in Hamiltonians with the time-reversal symmetry $\mathcal{T}$ and the spatial inversion symmetry $\mathcal{I}$, which can be parameterized in terms of three real numbers $K_1$, $K_2$, and $K_3$ as follows: 
\begin{equation}\label{eq:H123}
\mathcal{H}=K_{1}\mathcal{H}_{1}+K_{2}\mathcal{H}_{2}+K_{3}\mathcal{H}_{3}.
\end{equation}
Note that the model given in Eq.~\eqref{eq:H123} respects the $SU(2)_{\gamma}\times U(1)$ symmetry rather than the $SU(2)_{\gamma}$ symmetry.
For simplicity, we shall consider bipartite lattices only, including a 1D chain, a square lattice and a cubic lattice.

\subsection{Ground states}
\label{sec:gs}
\begin{figure}[htpb]
\centering
\includegraphics[width=8.4cm]{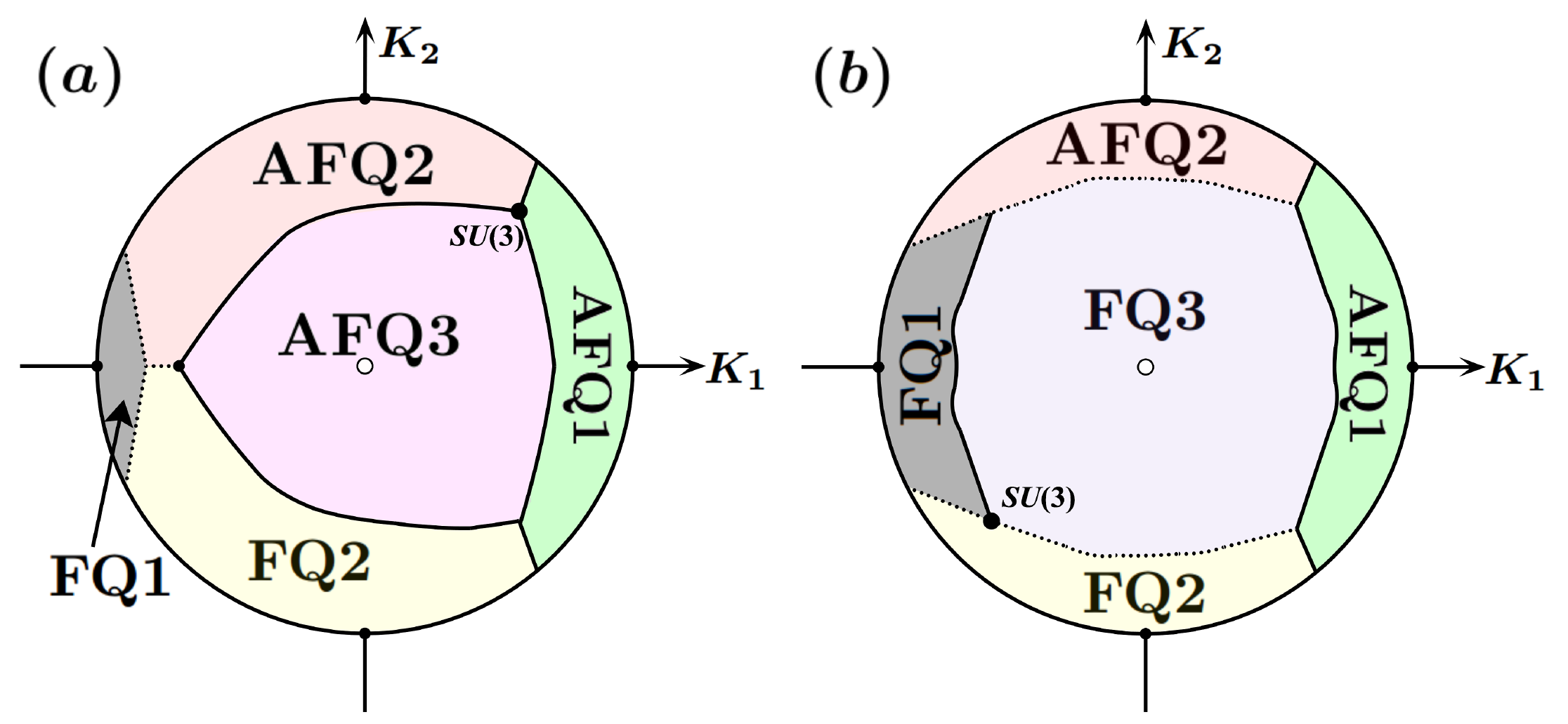}
\caption{Mean-field phase diagrams for $SU(2)\times U(1)\times\mathcal{T}\times\mathcal{I}$-symmetric models as defined in Eq.~\eqref{eq:H123} on bipartite lattices.
Two $SU(3)$ points are located at $K_1=K_2=K_3$. (a) Top view of the parameter space ($K_3>0$). (b) Bottom view of the parameter space ($K_3<0$).
Solid lines indicate first-order phase transitions, while dashed lines indicate continuous transitions. 
}\label{Fig:phases}
\end{figure}

To explore the ground-state phase diagram, we set $K^{2}_1+K^{2}_2+K^{2}_3=1$, such that the parameter space is a sphere.
Top and bottom views (along the $K_3$ axis) of this sphere are displayed in Fig.~\ref{Fig:phases}, 
where the mean-field phase diagram is presented.
There are six ordered phases, FQ1, FQ2, FQ3, AFQ1, AFQ2, and AFQ3. Here, FQ refers to a ferro-quadrupolar state, and AFQ refers to an antiferro-quadrupolar state (or, to be precise, a state with a staggered quadrupolar order).
When $K_{1(2,3)}$ is negative and predominates, the ground states are FQ states, while when $K_{1(2,3)}$ is positive and predominates, the ground states are AFQ states.
The solid lines in the phase diagram represent first-order transitions, while the dashed lines represent continuous transitions.
The $SU(3)$ symmetry will be achieved at two points where $K_1=K_2=K_3$. Both $SU(3)$ points are tricritical points. The one with $K_{1,2,3}<0$ corresponds to three phases, FQ1, FQ2 and FQ3, 
while the other one, with $K_{1,2,3}>0$, corresponds to AFQ1, AFQ2 and AFQ3.

\begin{figure}[htpb]
	\includegraphics[width=8.4cm]{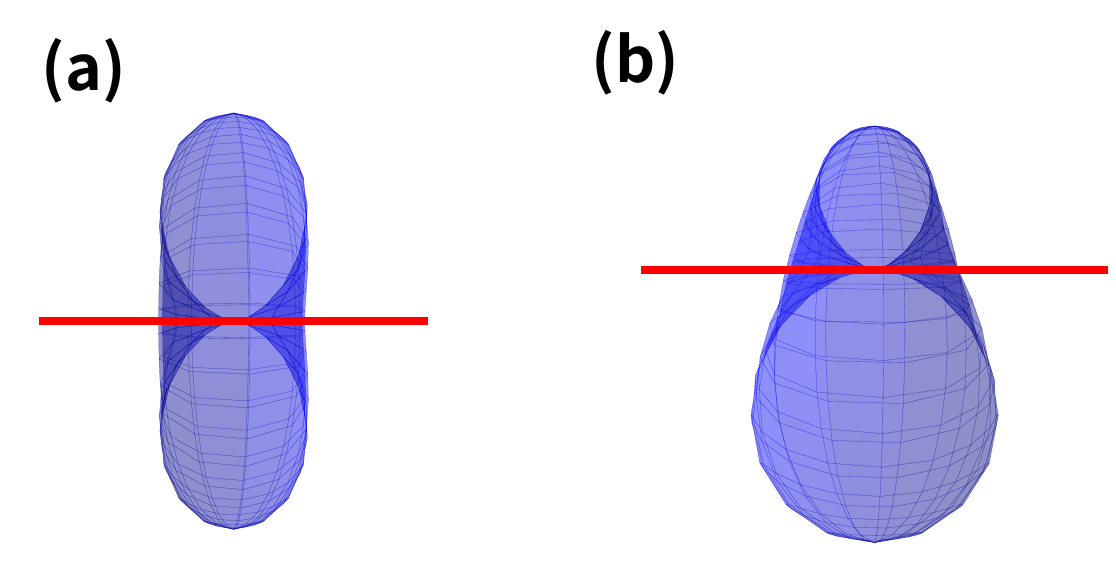}
	\caption{Local spin density for (a) time reversal invariant  and (b) time reversal breaking states. Red bars indicate the directions of $(|d_{x}|, |d_{y}|, |d_{z}|)$. Blue surfaces represent the spin density of a local wavefunction $\Psi$, which is defined as $|\langle\hat{S}(\hat{n})|\Psi\rangle|^{2}$. Here $|\hat{S}(\hat{n})\rangle$ is the spin coherent state pointing along direction $\hat{n}$ and is defined by $(\hat{n}\cdot\bm{S})|\hat{S}(\hat{n})\rangle=|\hat{S}(\hat{n})\rangle$. The local states $|\Psi\rangle$ are chosen for (a) $|\Psi\rangle=(|x\rangle+|y\rangle)/\sqrt{2}$; (b) $|\Psi\rangle=(e^{i\pi/10}|x\rangle+|y\rangle)/\sqrt{2}$. }\label{fig:wf0}
\end{figure}

We use local spin density to illustrate the wavefunctions for various FQ and AFQ states. 
If $\bm{d}$ vector is real, the state is time reversal invariant, and the local spin density $|\langle\hat{S}(\hat{n})|\bm{d}\rangle|^{2}$ is invariant under an $SO(2)$ rotation along the axis of $(|d_{x}|, |d_{y}|, |d_{z}|)$, which is so-called $SO(2)$ symmetric pattern. 
Otherwise, $|\bm{d}\rangle$ breaks the time-reversal symmetry, $|\langle\hat{S}(\hat{n})|\bm{d}\rangle|^{2}$ will be distorted from an $SO(2)$ symmetric pattern to an $SO(2)$ non-symmetric pattern. Two examples for time reversal invariant state and time reversal breaking state are given in FIG.~\ref{fig:wf0}

\begin{figure}[htpb]
	\includegraphics[width=8.4cm]{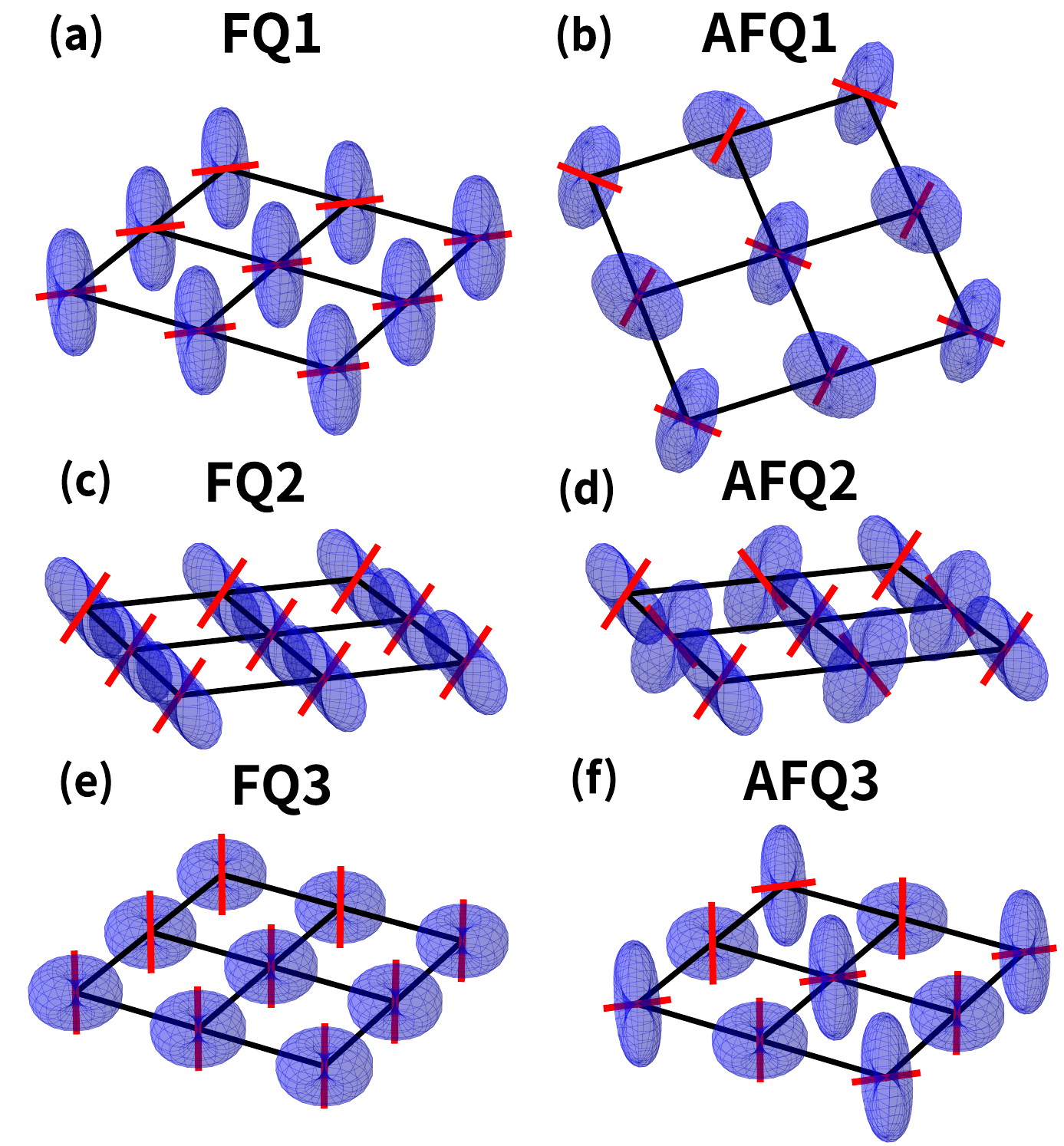}
	\caption{ Six types of spin quadrupolar orders on square lattice. All the states are time-reversal invariant with real $\bm{d}$ vectors, and red bars indicate the directions of these real $\bm{d}$ vectors. Blue surfaces represent the spin density of a local wavefunction $\Psi$, which is defined as $|\langle\hat{S}(\hat{n})|\Psi\rangle|^{2}$. Here $|\hat{S}(\hat{n})\rangle$ is the spin coherent state pointing along direction $\hat{n}$ and is defined by $(\hat{n}\cdot\bm{S})|\hat{S}(\hat{n})\rangle=|\hat{S}(\hat{n})\rangle$. The local states $|\Psi\rangle$ are chosen for FQ and AFQ phases as follows, (a) FQ1: $|\Psi\rangle=(|x\rangle+|y\rangle)/\sqrt{2}$; (b) AFQ1: $|\Psi_1\rangle=(|x\rangle+|y\rangle)/\sqrt{2}$ and $|\Psi_2\rangle=(|x\rangle-|y\rangle)/\sqrt{2}$; (c) FQ2: $|\Psi\rangle=(|x\rangle+|z\rangle)/\sqrt{2}$; (d) AFQ2: $|\Psi_1\rangle=(|x\rangle+|z\rangle)/\sqrt{2}$ and $|\Psi_2\rangle=(|x\rangle-|z\rangle)/\sqrt{2}$; (e) FQ3: $|\Psi\rangle=|z\rangle$; (f) AFQ3: $|\Psi_1\rangle=(|x\rangle+|y\rangle)/\sqrt{2}$ and $|\Psi_2\rangle=|z\rangle$. Here the subscripts in $\Psi_{1,2}$ refer to sublattices 1 and 2.
	}\label{fig:wf}
\end{figure}

All the six quadrupolar ordered states are illustrated on square lattice in FIG.~\ref{fig:wf}, where we choose time-reversal invariant ground states to eliminate spin dipolar orders and manifest quadrupolar orders. Notably, dipolar and quadrupolar orders may coexist in a ground state in the FQ1, FQ2, AFQ1 and AFQ2 phases, while only a quadrupolar order exists in the FQ3 and AFQ3 phases.

\subsection{Low energy excitations}
\label{sec:exci}
The low-energy excitations can be understood in the framework of the symmetry hierarchy as follows.
(1) The spontaneous symmetry breaking is distinct in the different phases: (a) $SU(2)$ is broken in FQ1 (AFQ1), but $U(1)$ is not (i.e., $SU(2)\times U(1) \to U(1)$); (b) both $SU(2)$ and $U(1)$ are broken in FQ2 (AFQ2) (i.e., $SU(2)\times U(1) \to 1$); and (c) neither $SU(2)$ nor $U(1)$ is broken in FQ3 (AFQ3).
(2) For FQ1, the $\omega^{\mbox{\tiny{FQ1}}}_{1}$ mode is gapless, while the other mode, $\omega^{\mbox{\tiny{FQ1}}}_{2}$, is gapful.
Since $SU(2)$ is broken, the gapless Goldstone mode $\omega^{\mbox{\tiny{FQ1}}}_{1}$ tends to recover the symmetry.
However, $U(1)$ is unbroken, so $\omega^{\mbox{\tiny{FQ1}}}_{2}$ is not required to be gapless as well. The gapless $\omega^{\mbox{\tiny{FQ1}}}_{1}$ mode corresponds to two-magnon excitations, while the gapful $\omega^{\mbox{\tiny{FQ1}}}_{2}$ mode corresponds to one-magnon excitations (see Appendix.\ref{app:FW}).  
(3) For FQ2, there are two gapless Goldstone modes, $\omega^{\mbox{\tiny{FQ2}}}_{1}$ and $\omega^{\mbox{\tiny{FQ2}}}_{2}$, because both $SU(2)$ and $U(1)$ are broken. The $\omega^{\mbox{\tiny{FQ2}}}_{1}$ mode is an admixture of one- and two-magnon excitations, while the $\omega^{\mbox{\tiny{FQ2}}}_{2}$ mode consists of one-magnon excitations only.
(4) For FQ3, there are two gapful modes, $\omega^{\mbox{\tiny{FQ3}}}_{1}=\omega^{\mbox{\tiny{FQ3}}}_{2}$, which are related to each other through the $SU(2)$ symmetry. Both of them correspond to one-magnon excitations.
(5) The AFQ1, AFQ2 and AFQ3 phases can be analyzed similarly.

The mean-field ground states and low-energy flavor-wave excitations for these six phases are summarized in Table \ref{tab:summarysix}. 

\begin{table*}[htbp]
	\renewcommand\arraystretch{1.3}
	\addtolength{\tabcolsep}{1.0ex}
	\begin{center}
		\caption{Summary of the $SU(2)_{\gamma}\times U(1) \times \mathcal{T} \times \mathcal{I}$-symmetric model defined in Eq.~\eqref{eq:H123}. The parameters $\vartheta$ and $\tilde{\vartheta}$ are given by
$\sin^{2}\vartheta=\frac{2|K_{2}|+K_{1}+K_{3}}{4|K_{2}|+K_{1}+3K_{3}}$ and $\sin^{2}\tilde{\vartheta}=\frac{2K_{2}+K_{1}+K_{3}}{4K_{2}+K_{1}+3K_{3}}$, respectively. $\mathcal{R}(\chi,\theta,\phi,\varphi)=\mbox{diag} (
e^{i\frac{\phi}{2}\sigma^{z}}e^{i\frac{\theta}{2}\sigma^{y}}e^{i\frac{\chi}{2}\sigma^{z}}, e^{i\varphi})$ is an $SU(2)\times U(1)$ rotation. $A_{K}$, $B_{K}$, $C_{K}$, $D_{K}$ and $\gamma(\bm{k})$ are defined as follows: 
$A_{K} = \frac{2|K_{2}|-K_{1}+K_{3}-2K_{1}K_{3}/|K_{2}|}{4|K_{2}|+K_{1}+3K_{3}}$,
$B_{K} = \frac{K_{1}+3K_{3}+2K_{3}(K_{1}+K_{3})/|K_{2}|}{4|K_{2}|+K_{1}+3K_{3}}$, 
$C_{K} = (3+K_{1}/K_{3})/2$,
$D_{K} = 2K_{2}/K_{3}$, and $\gamma(\bm{k}) = Z^{-1}\sum_{\delta}e^{i\bm{k}\cdot\bm{\delta}}$, where $Z=2D$ is the coordination number and $\bm{\delta}$ is a nearest-neighbor displacement.
``1" refers to one-magnon excitations, ``2" refers to two-magnon excitations, and ``1+2" refers to an admixture of one- and two-magnon excitations.
		 }\label{tab:summarysix}
	\end{center}
	\begin{tabular}{c|c|c|c|c}
		\hline\hline
	       & $\bm{d}$ vector(s)  &  Flavor-wave dispersion & Gap & Magnon \\
		\hline
    FQ1   & $\begin{array}{lll}\bm{d}&=&\mathcal{R}(\chi,\theta,\phi,0)\left(1,0,0\right)^T\end{array}$ 
    & $\begin{array}{lll}\omega^{\mbox{\tiny{FQ1}}}_{1}(\bm{k}) &=& 2Z|K_{1}|[1-\gamma(\bm{k})]\\ 
      \omega^{\mbox{\tiny{FQ1}}}_{2}(\bm{k}) &=& Z(|K_{1}|-K_{3})+2ZK_{2}\gamma(\bm{k})\end{array}$ 
    & $\begin{array}{c}\mbox{gapless}\\ \mbox{gapful}\end{array}$ & $\begin{array}{c}2\\1\end{array}$\\
		\hline
	 FQ2  & $\begin{array}{lll}\bm{d}&=&\mathcal{R}(\chi,\theta,\phi,\varphi)\left(\cos\vartheta,0,\sin\vartheta\right)^T\end{array}$ 
	 & $\begin{array}{lll} \omega^{\mbox{\tiny{FQ2}}}_{1}(\bm{k}) &=& 2Z|K_{2}|A_{K}(1-\gamma(\bm{k}))\\
      \omega^{\mbox{\tiny{FQ2}}}_{2}(\bm{k}) &=& 2Z|K_{2}|\sqrt{[1-\gamma(\bm{k})][1+B_{K}\gamma(\bm{k})]} \end{array}$ 
    & gapless & $\begin{array}{c}1+2\\1\end{array}$\\
	  \hline
	 FQ3  & $\begin{array}{lll}\bm{d}&=&\mathcal{R}(0,0,0,\varphi_{i})\left(0,0,1\right)^T\end{array}$ 
	 & $\begin{array}{lllll}\omega^{\mbox{\tiny{FQ3}}}_{1}(\bm{k}) & = & \omega^{\mbox{\tiny{FQ3}}}_{2}(\bm{k}) & = & 2Z[|K_{3}|+K_{2}\gamma(\bm{k})] \end{array}$ & gapful & $1$\\
		\hline
    AFQ1 & $\begin{array}{lll}\bm{d}_{1} & = & \mathcal{R}(\chi,\theta,\phi,0)\left(1,0,0\right)^T \\ \bm{d}_{2} & = & \mathcal{R}(\chi,\theta,\phi,0)\left(0,1,0\right)^T \end{array}$ 
	  & $\begin{array}{lllll} \omega^{\mbox{\tiny{AFQ1}}}_{1}(\bm{k}) & = & \omega^{\mbox{\tiny{AFQ1}}}_{2}(\bm{k}) & = & 2ZK_{1}\sqrt{1-\gamma^{2}(\bm{k})}\\
      \omega^{\mbox{\tiny{AFQ1}}}_{3}(\bm{k}) & = & \omega^{\mbox{\tiny{AFQ1}}}_{4}(\bm{k}) & = & Z(K_{1}-K_{3})\end{array}$ 
     & $\begin{array}{c}\mbox{gapless}\\ \mbox{gapful}\end{array}$ & $\begin{array}{c}2\\1\end{array}$\\
		\hline
		AFQ2  & $\begin{array}{lll}\bm{d}_{1} & = & \mathcal{R}(\chi,\theta,\phi,\varphi)\left(\cos\tilde{\vartheta},0,\sin\tilde{\vartheta}\right)^T \\ \bm{d}_{2} & = & \mathcal{R}(\chi,\theta,\phi,\varphi)\left(\cos\tilde{\vartheta},0,-\sin\tilde{\vartheta}\right)^T \end{array}$  
		& $\begin{array}{lll} \omega_{1}^{\mbox{\tiny{AFQ2}}} &=& 2ZK_{2}A_{K}[1-\gamma(\bm{k})]\\
      \omega_{2}^{\mbox{\tiny{AFQ2}}} &=& 2ZK_{2}A_{K}[1+\gamma(\bm{k})]\\
      \omega_{3}^{\mbox{\tiny{AFQ2}}} &=& 2ZK_{2}\sqrt{[1+\gamma(\bm{k})][1-B_{K}\gamma(\bm{k})]}\\
      \omega_{4}^{\mbox{\tiny{AFQ2}}} &=& 2ZK_{2}\sqrt{[1-\gamma(\bm{k})][1+B_{K}\gamma(\bm{k})]}\end{array}$ 
     & gapless & $\begin{array}{c}1+2\\1+2\\1\\1\end{array}$\\
		\hline
		AFQ3  & $\begin{array}{lll}\bm{d}_{1} & = & \mathcal{R}(0,0,0,\varphi_{i})\left(0,0,1\right)^T \\ \bm{d}_{2} & = & \mathcal{R}(\chi_{i},\theta_{i},\phi_{i},0)\left(1,0,0\right)^T \end{array}$ 
		& $\begin{array}{lll} \omega^{\mbox{\tiny{AFQ3}}}_{1}(\bm{k}) &=& Z(K_{3}-K_{1})\\
      \omega^{\mbox{\tiny{AFQ3}}}_{2}(\bm{k}) &=& 0+O(\bm{k}^{3})\\
      \omega^{\mbox{\tiny{AFQ3}}}_{3}(\bm{k}) &=& Z[K_{3}\sqrt{C_{K}^{2}-D_{K}^{2}\gamma^{2}(\bm{k})}+\frac{K_{1}-K_{3}}{2}]\\
      \omega^{\mbox{\tiny{AFQ3}}}_{4}(\bm{k}) &=& Z[K_{3}\sqrt{C_{K}^{2}-D_{K}^{2}\gamma^{2}(\bm{k})}+\frac{K_{3}-K_{1}}{2}]\end{array}$ & $\begin{array}{c}\mbox{gapful}\\ \mbox{gapless} \\ \mbox{gapful} \\ \mbox{gapful}\end{array}$ & $\begin{array}{c}2\\2\\1\\1\end{array}$ \\
		\hline\hline
	\end{tabular}
\end{table*}

\subsection{Spectral functions: A clue to the eight-fold way}
\label{sec:spec}
Inelastic neutron scattering measures the spin spectral function in $(\bm{q},\omega)$ space, which is defined as $$S^{\alpha\beta}(\bm{q},\omega)=\mbox{Im}\left\{i\int_{0}^{\infty}dt{}e^{i\omega{}t}\langle[S ^{\alpha}(\bm{q},t),S^{\beta}(-\bm{q},0)]\rangle\right\},$$
where we have set $\mu{}g_{B}=1$ for simplicity. 
At zero temperature, $S^{\alpha\beta}(\bm{q},\omega)$ depends on the choice of the ground state. However, the spectral function, $$S(\bm{q},\omega)\equiv S^{xx}(\bm{q},\omega)+S^{yy}(\bm{q},\omega)+S^{zz}(\bm{q},\omega),$$
does not change qualitatively within a single phase.
On the other hand, resonant inelastic X-ray scattering (RIXS) measures two-magnon processes, which is described by spin quadrupole spectral functions, $$Q^{\mu\nu}(\bm{q},\omega)=\mbox{Im}\left\{i\int_{0}^{\infty}dt{}e^{i\omega{}t}\langle[Q ^{\mu}(\bm{q},t),Q^{\nu}(-\bm{q},0)]\rangle\right\},$$ and $\mu$ and $\nu$ denote $xy,\ yz,\ zx,\ x^{2}-y^{2}$ and $3z^{2}-r^{2}$. Similar to $S(\bm{q},\omega)$, the spin quadrupole spectral function, $$Q(\bm{q},\omega)\equiv \sum_{\mu}Q^{\mu\mu}(\bm{q}, \omega),$$ does not change qualitatively within a single phase as well.
Therefore, the spin spectral function $S(\bm{q},\omega)$ and the spin quadrupole spectral function $Q(\bm{q},\omega)$ can be used to detect flavor waves and distinguish the various FQ and AFQ phases. 

In the flavor-wave mean-field theory, all the degenerate ground states can be obtained from one of them by an $SU(2)\times U(1)$ rotation of the $\bm{d}$-vector. We parameterize a general $SU(2)$ rotational matrix as 
\begin{equation}\label{eq:R_ri}
R = r_{0}\sigma^{0} + i\sum_{n=1}^{3}r_{n}\sigma^{n},
\end{equation}
where $\sigma^{0}$ are identity as well as $\sigma^{1}$, $\sigma^{2}$ and $\sigma^{3}$ are three Pauli matrices. Here $\hat{r}=\{r_{0},r_{1},r_{2},r_{3}\}$ is a 4-dimensional real vector with $\sum_{n=0}^{3}r^{2}=1$. Thus, apart from a global phasefactor, two $\bm{d}$ vectors of two degenerate ground states, $\bm{d}$ and $\bm{d}^{\prime}$, are related by a $3\times 3$ matrix as follows, 
\begin{equation}
\bm{d}^{\prime} = \left(\begin{array}{cc}
e^{-i\phi}R & 0_{2,1}\\
0_{1,2} & 1 \\
\end{array}\right)\bm{d},
\end{equation}
where $0_{1,2}$($0_{2,1}$) is a $1\times{}2$($2\times{}1$) zero matrix. In the $SU(3)$ Schwinger boson representation, the expression for the spin operators $\bm{S}$ depends on the parameters $r_{0-3}$.

\begin{figure*}[htbp]
	\centering
	\includegraphics[scale=0.62]{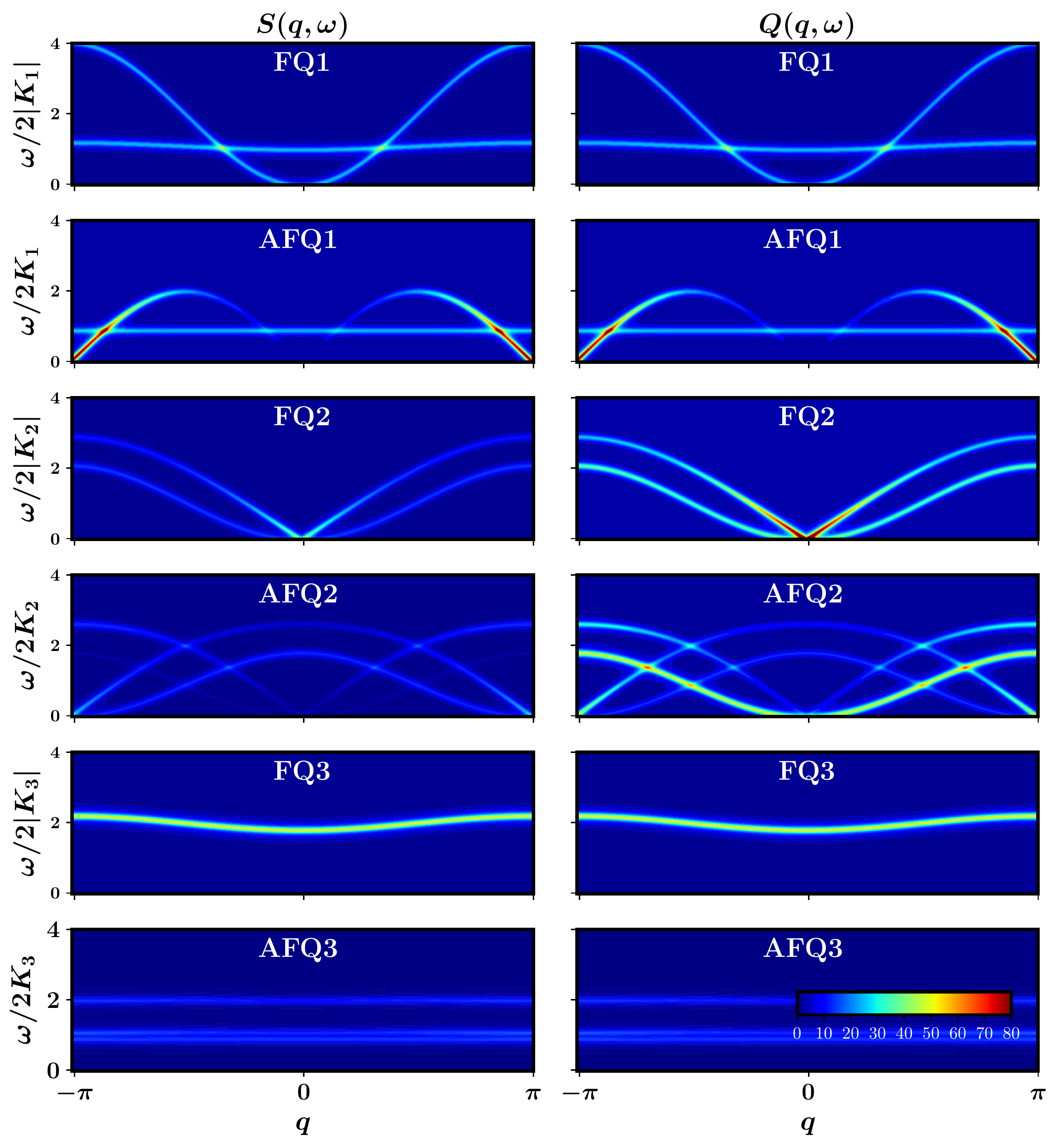}
	\caption{The spin spectral functions $S(\bm{q},\omega)$ and the spin quadrupole spectral functions $Q(\bm{q},\omega)$ for the FQ and AFQ phases. Here, we set $K_{2}=K_{3}=0.1K_{1}$ for FQ1 and AFQ1, $K_{1}=K_{3}=0.1K_{2}$ for FQ2 and AFQ2 and $K_{1}=K_{2}=0.1K_{3}$ for FQ3 and AFQ3.}\label{fig:Sqqw}
\end{figure*}

Thus these spectral functions can be evaluated for each FQ or AFQ state; these functions are distinct in different phases but do not qualitatively change within a single phase. 
Moreover, $S(\bm{q},\omega)$ and $Q(\bm{q},\omega)$ {\em share the same structure} as long as the elementary excitations are flavor waves, as demonstrated in Fig.~\ref{fig:Sqqw}.
Namely, $Q(\bm{q},\omega)$ has the same dispersion as $S(\bm{q},\omega)$, and difference between them is in the spectral weight. {\em This similarity provides evidence for the underlying $SU(3)$ structure and serves as a clue to the eight-fold way.} 
The details of these spectral functions for all FQ and AFQ phases can be found in see Appendix.\ref{app:sf}.

\section{Beyond the mean-field theory}\label{sec:beyondMFT}
\subsection{Effective Hamiltonian in the AFQ3 phase: A possible emergent $S=1/2$ gapless spin liquid}
\label{sec:effqsl}
\begin{figure}[htpb]
	\centering
	\includegraphics[scale=0.23]{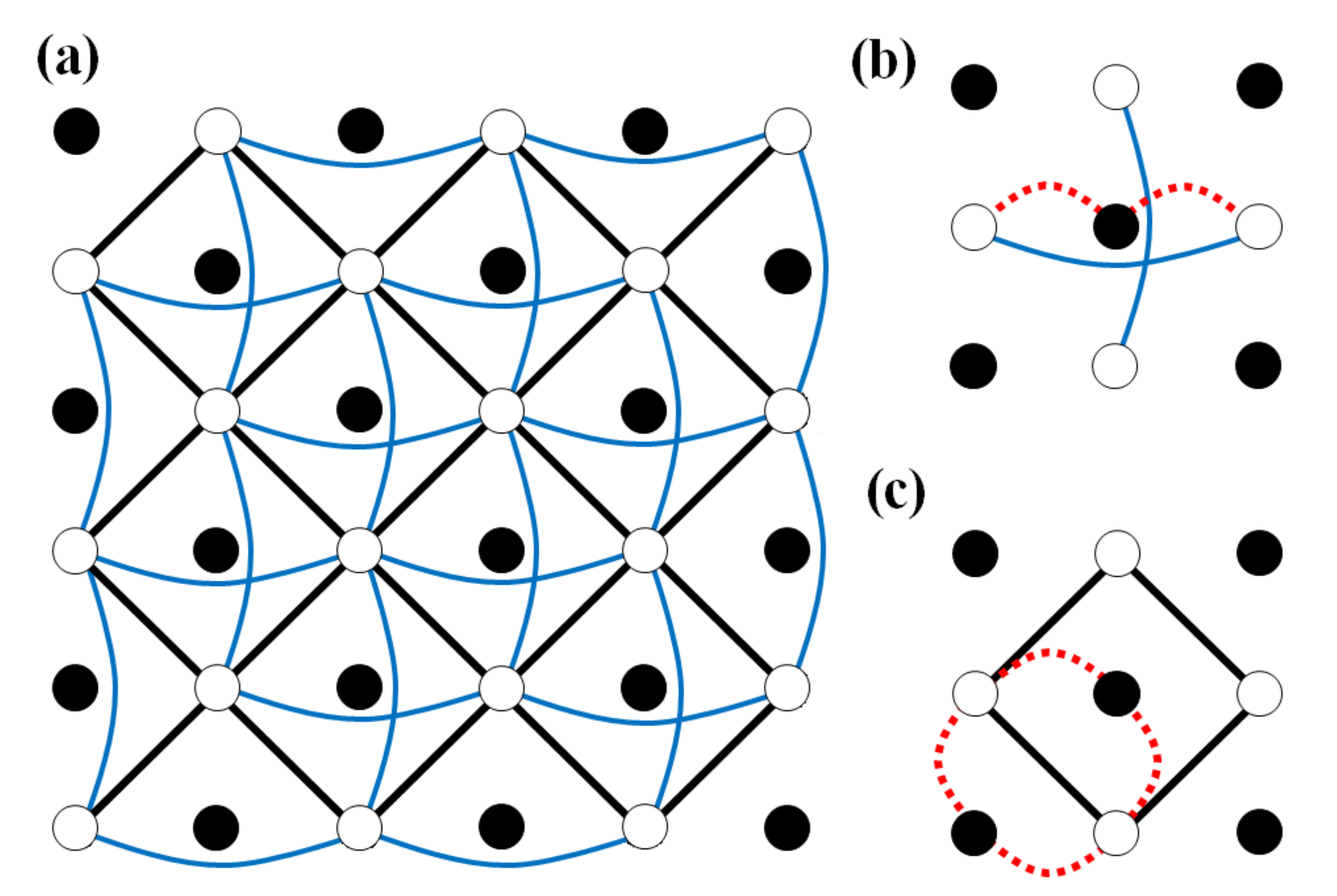}
	\caption{Solid circles form sublattice-$1$ where all the spins are in the $|z\rangle$ state and open circles form sublattice-$2$ where all the spins are in $|x\rangle$ or $|y\rangle$ state. (a) Emergent $\tilde{J}_{1}-\tilde{J}_{2}$ super square lattice with black solid lines for $\tilde{J}_{1}$ bonds and blue curve lines for $\tilde{J}_{2}$ bonds. (b) one path to achieve $J_{2}$ bond. (c) two paths to achieve $\tilde{J}_{1}$ bonds. (b) and (c) indicate that $\tilde{J}_{2}/\tilde{J}_{1}=0.5$.
	}\label{fig:heff}
\end{figure}

In the mean-field solution, the AFQ3 ground states are locally degenerate inside a bulk energy gap. 
This huge degeneracy arises from the unperturbative Hamiltonian $K_3\mathcal{H}_{3}$, of which ground state subspace is spanned by the local basis $\{|x\rangle_i,|y\rangle_i\}$ on sublattice-2 and $\{|z\rangle_i\}$ on sublattice-1, as shown in FIG.\ref{fig:heff}(a).
And the degeneracy is expected to be lifted by a small but finite $K_1$ and $K_2$. 
To address this case and go beyond the mean-field theory, we consider perturbations of up to the third order in the limit $K_3\gg |K_{1(2)}|$, where the unperturbative energy gap is about $ZK_3$.
What we need is to include all the possible perturbations of $\mathcal{H}_1$ and $\mathcal{H}_2$ and project the states into the subspace of unperturbative ground states by a projector $\mathcal{P}$.

We begin with two-site Hamitonians $h_{i}$, $i=1,2,3$, which is the simplest case of the Hamiltonians $\mathcal{H}_{i}$ defined in Eq.~\eqref{eq:H1-6}, and consider their matrix elements. Firstly, $h_{3} = Q^{3z^{2}-r^{2}}_{1}Q^{3z^{2}-r^{2}}_{2}$, which is diagonal in the basis $\{|\sigma_1\sigma_2\rangle\}$, where $\sigma_{1,2}=x,y,z$. 
Secondly, $h_{1}=S^{z}_{1}S^{z}_{2}+Q^{xy}_{1}Q^{xy}_{2}+Q^{x^{2}-y^{2}}_{1}Q^{x^{2}-y^{2}}_{2}$, and we have
\begin{equation}\label{eq:h1}
h_{1}|\sigma_{1}\sigma_{2}\rangle = \left\lbrace \begin{array}{cl}
|\sigma_{1}\sigma_{2}\rangle, & \sigma_{1}=\sigma_{2} \neq{}z, \\
2|\sigma_{2}\sigma_{1}\rangle, & \sigma_{1}=x(y),\sigma_{2}=y(x), \\
0, & \mbox{otherwise}.
\end{array}\right.
\end{equation}
Thirdly, $h_{2}={}S^{x}_{1}S^{x}_{2}+S^{y}_{1}S^{y}_{2}+Q^{yz}_{1}Q^{yz}_{2}+Q^{zx}_{1}Q^{zx}_{2}$, we have
\begin{equation}\label{eq:h2}
h_{2}|\sigma_{1}\sigma_{2}\rangle = \left\lbrace \begin{array}{cl}
2|\sigma_{2}\sigma_{1}\rangle, & \sigma_{1} =x,y(z)\  \mbox{and}\  \sigma_{2}=z(x,y), \\
0, & \mbox{otherwise}.
\end{array}\right.
\end{equation}
Taking into account all the nonzero matrix elements, the leading perturbation is of the third order and the effective Hamiltonian reads
\begin{equation}
\mathcal{H}_{eff} =	\frac{K_{1}K^{2}_{2}}{K_{3}^{2}} \mathcal{P}\mathcal{H}_{2}\mathcal{H}_{1}\mathcal{H}_{2}\mathcal{P}.
\end{equation}
Regarding the projector $\mathcal{P}$, $\mathcal{H}_{eff}$ can be written as a $\mathcal{H}_{1}$ type Hamiltonian defined on one of the sublattices.

As an example, consider a square lattice; the spins have a quadrupolar order on one of the two sublattices, and we have the following effective Hamiltonian on the other sublattice:
\begin{eqnarray}
\mathcal{H}_{eff} & = & \tilde{J}_1 \sum_{\langle ij\rangle_1} \mathcal{P}\left(S^{z}_{i}S^{z}_{j}+Q^{xy}_{i}Q^{xy}_{j}+Q^{x^{2}-y^{2}}_{i}Q^{x^{2}-y^{2}}_{j}\right)\mathcal{P} \nonumber \\
&& + \tilde{J}_2 \sum_{\langle ij\rangle_2}\mathcal{P}\left(S^{z}_{i}S^{z}_{j}+Q^{xy}_{i}Q^{xy}_{j}+Q^{x^{2}-y^{2}}_{i}Q^{x^{2}-y^{2}}_{j}\right)\mathcal{P}\nonumber,
\end{eqnarray}
where $\langle ij\rangle_{1(2)}$ denotes a pair of (next) nearest neighboring sites on the sublattice, $\tilde{J}_1=2K_{1}K^{2}_{2}/K_{3}^{2}$ and $\tilde{J}_2=K_{1}K^{2}_{2}/K_{3}^{2}$.
Note that the relation $\tilde{J}_2/\tilde{J}_1=1/2$ is guaranteed by the fact that there are two paths contribute to $\tilde{J}_1$ while only one path contributes to $\tilde{J}_2$, as illustrated in FIG.~\ref{fig:heff}

Thus, there emerges an effective spin-$1/2$ $\tilde{J}_1$-$\tilde{J}_2$ Heisenberg model constructed by the $SU(2)_{\gamma}$ generators.
When $K_1<0$, $\tilde{J}_{1(2)}<0$, and the ground state is of ferromagnetic order. When $K_1>0$, $\tilde{J}_1=2\tilde{J}_2>0$, a gapless quantum spin liquid (QSL) ground state~\cite{RMPQSL} may be favored, as suggested by variational Monte Carlo~\cite{j1j2_gapless_vmc, j1j2_gapless_vmc2, j1j2_gapless_vmc3} and density matrix renormalization group (DMRG)~\cite{j1j2_gapless_dmrg}. It is worth noting that the ground states of spin-$1/2$ $J_1$-$J_2$ Heisenberg model still remains a controversial issue, and various numerical approaches lead to several contradictory results. Other possible ground states include N\'{e}el order\cite{j1j2_neel_peps, j1j2_neel_peps2} (by PEPS), gapful spin liquid~\cite{j1j2_gapped_dmrg} (by DMRG), and plaquette valence-bond solid.\cite{j1j2_PVB_ed, j1j2_PVB_ed2, j1j2_PVB_trg, j1j2_PVB_dmrg} (by exact diagonalization, tensor network, and DMRG). Despite of the controversy, we would like to propose that a QSL state with effective spin $S=1/2$ (or other ordered states) may coexist with gapful quadrupolar order in a quantum spin-1 system.


\subsection{Hydrodynamics modes in a quantum spin-orbital liquid: In analogy to QCD}
\label{sec:hydro}
Now we consider a quantum spin-orbital liquid ground state in the $SU(2)\times U(1)$ symmetric model, where neither $SU(2)$ nor $U(1)$ symmetry is broken and all the correlation functions of the operators $\{\bm{S},\bm{Q}\}$ are short ranged. If there does exist such a spin-orbital liquid state, the low energy excitations can be described by the fields of $\{\bm{S},\bm{Q}\}$ and can be classified according to the $SU(2)\times U(1)$ symmetry. Because of the short ranged correlation, these excitations are gapful. In analogy to QCD, we will demonstrate below that these excitation gaps must satisfy some relations due to the symmetry hierarchy $SU(3)\supset SU(2)\times U(1)$. 

An interesting observation is that the three local spin states $|x\rangle,|y\rangle,|z\rangle$ can be naturally in analogy with the $u,d,s$ quarks in particle physics in the fundamental representation of $SU(3)$ symmetry as follows,
\begin{equation}\label{eq:uds}
\renewcommand\arraystretch{1.25}
\begin{array}{lll}
|x\rangle & \longleftrightarrow & u, \\
|y\rangle & \longleftrightarrow & d, \\
|z\rangle & \longleftrightarrow & s.
\end{array}
\end{equation}
The quark model is a successful theory of the strong interaction, which is known as QCD. According to Gell-Mann's argument: (1) There exists an additive quantum number called strangeness is conserved in addition to isospin $SU(2)$ symmetry; (2) in very strong interactions region, the symmetry is $SU(3)$ rather then the $SU(2)\times{}U(1)$; (3) in medium strong interactions region, the $SU(3)$ breaks into $SU(2)\times{}U(1)$, i.e., isospin $SU(2)$ and hypercharge $U(1)$. This symmetry hierarchy is exactly as the same as what we discuss in our spin-1 systems.
Moreover, the $SU(3)$ octet $\{\bm{S},\bm{Q}\}$ can be mapped to high energy particles, e.g., the light spin-0 mesons, in addition to the $SU(3)$ triplet mapping in Eq.~\eqref{eq:uds} as follows, 
\begin{equation}\label{eq:mesons}
\renewcommand\arraystretch{1.25}
\begin{array}{llll}	
\frac{1}{2}(iS^{x}-Q^{yz})& \longleftrightarrow & K^{0} &  (d\bar{s}), \\
\frac{1}{2}(-iS^{y}-Q^{zx})& \longleftrightarrow & K^{+} & (u\bar{s}), \\
\frac{1}{2}(iS^{x}+Q^{yz})& \longleftrightarrow & \bar{K}^{0} & (s\bar{d}), \\
\frac{1}{2}(-iS^{y}+Q^{zx})& \longleftrightarrow & K^{-} & (s\bar{u}), \\
\frac{1}{2}(iS^{z}-Q^{xy}) & \longleftrightarrow & \pi^{+} & (u\bar{d}), \\
\frac{1}{2}(-iS^{z}-Q^{xy}) & \longleftrightarrow & \pi^{-} & (d\bar{u}), \\
Q^{x^{2}-y^{2}}  & \longleftrightarrow & \pi^{0} & (u\bar{u},d\bar{d}), \\
Q^{3z^2-r^2}  & \longleftrightarrow & \eta & (u\bar{u},d\bar{d},s\bar{s}). 
\end{array}
\end{equation}

For the shorthand, let us define the following operators,
\begin{equation}\label{eq:hmodes}
\renewcommand\arraystretch{1.25}
\begin{array}{lll}
K^{0} &=& (iS^{x}-Q^{yz})/2,\\
K^{+} &=& (-iS^{y}-Q^{zx}),\\
\bar{K}^{0} &=& (-iS^{x}-Q^{yz})/2,\\
K^{-} &=& (iS^{y}-Q^{zx})/2,\\
\pi^{+} &=& (iS^{z}-Q^{xy})/2,\\
\pi^{-} &=& (-iS^{z}-Q^{xy})/2,\\
\pi^{0} &=& Q^{x^{2}-y^{2}},\\
\eta &=& Q^{3z^2-r^2}.
\end{array} 
\end{equation}
Thus, in the hydrodynamic limit, $\bm{q}\to 0$ and $\omega \to 0$, the collective modes in a quantum spin-orbital liquid can be described by the fields of $K^{0}$, $\bar{K}^{0}$, $K^{\pm}$, $\pi^{0}$, $\pi^{\pm}$ and $\eta$. 
And we expect that each hydrodynamic mode has an excitation gap $M$. The excitaiton gap is nothing but the mass of the corresponding particle.
By the $SU(2)\times U(1)$ symmetry, we deduce that these gaps must satisfy the relations, 
\begin{subequations}
\begin{eqnarray}
M_{K^{0}} & = & M_{K^{+}}, \\
M_{\bar{K}^{0}} & = & M_{K^{-}}, \\
M_{\pi^{\pm}} & = & M_{\pi^{0}} = M_{\pi},
\end{eqnarray}
as well as the famous Gell-Mann-Okubo formula \cite{Georgi},
\begin{equation}
2(M_{K^{+}}+M_{K^{-}})=3M_{\eta}+M_{\pi}.
\end{equation}
\end{subequations}

\section{Summary}
\label{sec:summary}
In summary, we have revealed hidden $SU(2)$ symmetries in spin-1 quantum magnets, studied them in accordance with the $SU(3)\supset SU(2)\times U(1)$ symmetry hierarchy, demonstrated novel emergent phenomena, and found some clues to the emergent eight-fold way. These $SU(2)$ symmetries may be realized in cold atoms as well as $d^8$ and/or $d^6$ electrons with the proper specific choices of the spin-orbital couplings.

\section{Acknowledgement}
We would like to thank Dong-Hui Xu, Hong-Hao Tu, Gang Chen, Hong Yao and Zheng-Yu Weng for helpful discussions.
This work is supported in part by National Key Research and Development Program of China (No.2016YFA0300202), National Natural Science Foundation of China (No.11774306), the Strategic Priority Research Program of Chinese Academy of Sciences (No. XDB28000000) and the Fundamental Research Funds for the Central Universities in China. 
JJM. is supported by China Postdoctoral Science Foundation(Grant No.2017M620880) and the National NaturalScience Foundation of China (Grant No.1184700424).

\appendix

\section{Fundamentals of $SU(3)$ Lie algebra}\label{app:SU3}
The eight Gell-Mann matrices are defined as,
\begin{equation}
\begin{split}
&\lambda_{1}=\left(\begin{array}{ccc}
0 & 1 & 0 \\
1 & 0 & 0 \\
0 & 0 & 0\\
\end{array}\right),\quad
\lambda_{2}=\left(\begin{array}{ccc}
0 & -i & 0 \\
i & 0 & 0 \\
0 & 0 & 0\\
\end{array}\right)\\
&\lambda_{3}=\left(\begin{array}{ccc}
1 & 0 & 0 \\
0 & -1 & 0 \\
0 & 0 & 0\\
\end{array}\right),\quad
\lambda_{4}=\left(\begin{array}{ccc}
0 & 0 & 1 \\
0 & 0 & 0 \\
1 & 0 & 0\\
\end{array}\right)\\
&\lambda_{5}=\left(\begin{array}{ccc}
0 & 0 & -i \\
0 & 0 & 0 \\
i & 0 & 0\\
\end{array}\right),\quad
\lambda_{6}=\left(\begin{array}{ccc}
0 & 0 & 0 \\
0 & 0 & 1 \\
0 & 1 & 0\\
\end{array}\right)\\      
&\lambda_{7}=\left(\begin{array}{ccc}
0 & 0 & 0 \\
0 & 0 & -i \\
0 & i & 0\\
\end{array}\right),\quad
\lambda_{8}=\frac{1}{\sqrt{3}}\left(\begin{array}{ccc}
1 & 0 & 0 \\
0 & 1 & 0 \\
0 & 0 & -2\\
\end{array}\right).\\
\end{split}
\end{equation}
The generators of $SU(3)$ Lie group are given by $T_i=\lambda_i/2$, $i=1,\cdots,8$.

In $SU(3)$ representations, a state in an irreducible representation (IR) is labelled by $(p,q)$, corresponding to the weight vector $\bm{\mu}=p\bm{\mu}^1+q\bm{\mu}^2$, where $\bm{\mu}^1=(1/2,\sqrt{3}/6)$ and $\bm{\mu}^2=(1/2,-\sqrt{3}/6)$.
The weights are defined by the eigenvalues of the Cartan generators $H_1$ and $H_2$, $H_i|\bm{\mu}\rangle=\mu_i|\bm{\mu}\rangle$ where $i=1,2$.
So that $T_3|(p,q)\rangle=(p+q)/2|(p,q)\rangle$ and $T_8|(p,q)\rangle=\sqrt{3}(p-q)/6|(p,q)\rangle$.
An IR is characterized by the highest weight $(n,m)$. Thus a state in a $SU(3)$ IR can be written as $|(n,m),(p,q)_k\rangle$. 
Note that there may exist more than one $(p,q)$ state in IR $(n,m)$, these different $(p,q)$ states are distinguished by the subscript $k$, which will be neglected when there is only one $(p,q)$ state.

\section{$SU(3)$ structure and Hidden $SU(2)$ symmetries}
\label{app:SU2}
Firstly, it is straightforward to examine the $SU(3)$ Lie algebra relation among $\{\bm{S},\bm{Q}\}$ through the commutators $[S^{\alpha},S^{\beta}]$, $[S^{\alpha},Q^{\mu}]$ and $[Q^{\mu},Q^{\nu}]$ directly. 
As mentioned, besides the $SO(3)$ subalgebra of $\{S^{x},S^{y},S^{z}\}$, there are other $SU(2)$ subalgebras belonging to the $SU(3)$ Lie algebra.

In order to find out the other $SU(2)$ subalgebras, we consider the Cartan subalgebra $H$, the largest commutitative subalgebra, of the $SU(3)$ Lie algebra,
which can be chosen to be made of linear combinations of two commutative operators $H_1=T_{3}$ and $H_2=T_{8}$, where $T_i=\lambda_i/2$
and satisfy $\mbox{Tr}(H_i H_j)=\frac{1}{2}\delta_{ij}$.
An $SU(2)$ subalgebra can be constructed as follows. Let us select an operator in the Cartan subalgebra $H$, which serves as $J^z$ in the $SU(2)$ algebra.  
Writing $\bm{H}=\{H_{1},H_{2}\}$, we have $J^z=|\bm{\alpha}|^{-2}\bm{\alpha}\cdot\bm{H}$, where $\bm{\alpha}$ is a two dimensional vector.
Then the raising and lowering operators $J^{\pm}$ can be obtained through $J^{\pm}=|\bm{\alpha}|^{-1}E_{\pm\bm{\alpha}}$, with $\bm{\alpha}$ a root vector.
It is easy to verify that $[J^z,J^{\pm}]=\pm J^{\pm}$ and $[J^+,J^{-}]=2J_z$.
So that a nonzero root $\bm{\alpha}$ of $SU(3)$ will give rise to an $SU(2)$ subgroup.

\begin{figure}[htpb]
	\includegraphics[width=7.6cm]{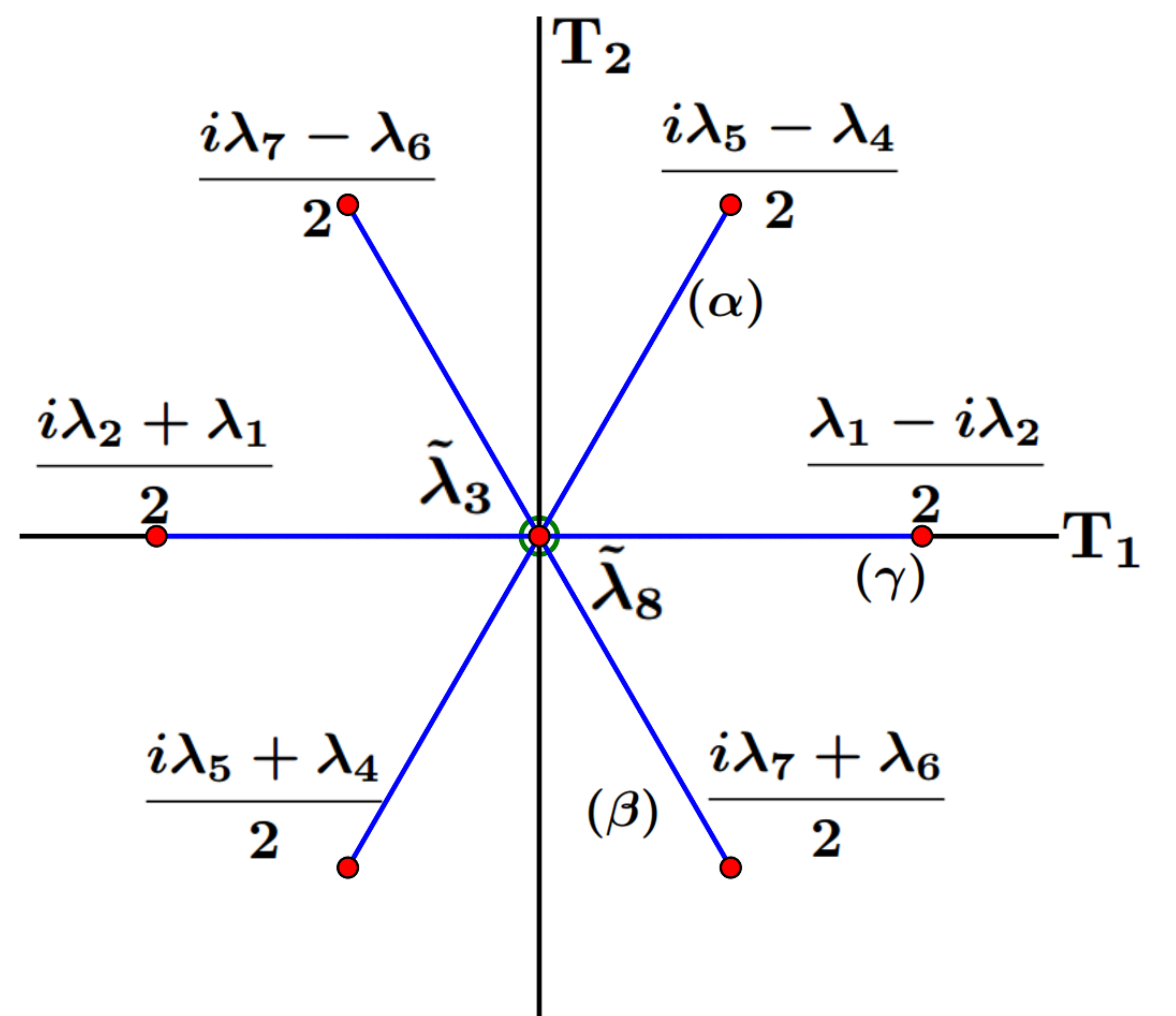}
	\caption{Roots of $SU(3)$ Lie algebra. There are two simple roots, $\bm{\alpha}=(\frac{1}{2},\frac{\sqrt{3}}{2})$ and $\bm{\beta}=(\frac{1}{2},-\frac{\sqrt{3}}{2})$. 
		$\bm{\gamma}=\bm{\alpha}+\bm{\beta}=(1,0)$ is the other positive root. $\{\frac{iT_2+T_1}{2},\frac{T_1-iT_2}{2},T_3\}$ are three generators of the subalgebra $SU(2)_{\gamma}$, and so on and so forth.
	}
	\label{Fig:IR11}
\end{figure}

The roots of $SU(3)$ algebra are nothing but the weights of its adjoint representation $(1,1)$, which are plotted in Fig.~\ref{Fig:IR11}. 
It is clear that there are three pairs of nonzero roots, $\{\pm\bm{\alpha},\pm\bm{\beta},\pm\bm{\gamma}\}$, where $\bm{\alpha}=(\frac{1}{2},\frac{\sqrt{3}}{2})$ and $\bm{\beta}=(\frac{1}{2},-\frac{\sqrt{3}}{2})$ are two simple roots, 
and $\bm{\gamma}=(1,0)$ is the other positive root with $\bm{\gamma}=\bm{\alpha}+\bm{\beta}$.  
So that $\bm{\alpha},\bm{\beta},\bm{\gamma}$ give rise to three $SU(2)$ subalgebras, whose generators are given as follows,
$SU(2)_{\alpha}:\{T_{4},T_{5},\bm{\alpha}\cdot\bm{H}\}$, $SU(2)_{\beta}:\{T_{6},T_{7},\bm{\beta}\cdot\bm{H}\}$ and $SU(2)_{\gamma}:\{T_{1},T_{2},\bm{\gamma}\cdot\bm{H}\}$. 
In terms of $\bm{S}$ and $\bm{Q}$, the generators of the three $SU(2)$ subgroups read,
\begin{eqnarray}
SU(2)_{\alpha}&:&\{Q^{zx},S^{y},\frac{1}{2}Q^{x^2-y^2}+\frac{\sqrt{3}}{2}Q^{3z^2-r^2}\},\nonumber\\
SU(2)_{\beta}&:&\{Q^{yz},S^{x},\frac{1}{2}Q^{x^2-y^2}-\frac{\sqrt{3}}{2}Q^{3z^2-r^2}\},\nonumber\\
SU(2)_{\gamma}&:&\{Q^{xy},S^{z},Q^{x^2-y^2}\}.\label{eq:SU2SQ}
\end{eqnarray}

The underlying $SU(3)$ structure and the hidden $SU(2)$ symmetries will be more transparent in the Cartesian coordinate representation of spin states, 
\begin{equation}\label{eq:xyz}
|x\rangle=i\frac{|1\rangle-|-1\rangle}{\sqrt{2}},\ |y\rangle=\frac{|1\rangle+|-1\rangle}{\sqrt{2}},\ |z\rangle=-i|0\rangle.
\end{equation}
It is easy to verify that $|\alpha\rangle$ is time reversal invariant and satisfy the relations $\langle \alpha |\beta\rangle =\delta_{\alpha\beta}$ and $S^{\alpha}|\beta\rangle=i\epsilon^{\alpha\beta\gamma}|\gamma\rangle$,
where $\alpha,\beta,\gamma = x,y,z$ and $\epsilon^{\alpha\beta\gamma}$ is the three-rank antisymmetric tensor.
Thus a spin state can be expressed as follows,
\begin{equation}\label{eq:d-vector}
|\bm{d}\rangle=d^{x}|x\rangle+d^{y}|y\rangle+d^{z}|z\rangle,
\end{equation}
where $\bm{d}=\left(d^{x},d^{y},d^{z}\right)$ is a complex vector, and normalization condition is given by $|\bm{d}|^{2}=1$.
So that a time reversal invariant state is given by a real vector $\bm{d}$ up to a global phase factor and characterized by $\langle \bm{d}|\bm{S}|\bm{d}\rangle=0$.
The expectation values for $\{\bm{S},\bm{Q}\}$ can be expressed in terms of the $\bm{d}$ vector,  $\langle S^{\alpha}\rangle=-i\epsilon_{\alpha\beta\gamma}\bar{d}^{\beta}d^{\gamma}$, 
$\langle Q^{\alpha\beta} \rangle|_{\alpha\neq\beta}=-(\bar{d}^{\alpha}d^{\beta}+\bar{d}^{\beta}d^{\alpha})$, $\langle Q^{x^{2}-y^{2}}\rangle=|d^{y}|^{2}-|d^{x}|^{2}$ and $\langle Q^{3z^{2}-r^{2}}\rangle=\frac{1}{\sqrt{3}}(2|d^{z}|^{2}-|d^{y}|^{2}-|d^{x}|^{2})$,
where $\bar{d}^{\alpha}$ is the conjugate complex number of $d^{\alpha}$. Then the path integral for a spin $S=1$ system can be written as in Eq.~\eqref{eq:pathint},
where the Hamiltonian $\mathcal{H}$ is given by Eq.~\eqref{eq:Hgeneric}, while the operators $\bm{S}$ and $\bm{Q}$ are replaced by their expectation values as follows,
\begin{equation}\label{eq:SQSU3d}
\begin{split}
& \left(\begin{array}{c}
S^{x} \\
S^{y} \\
S^{z} \\
Q^{x^{2}-y^{2}} \\
Q^{3z^{2}-r^{2}} \\
Q^{xy} \\
Q^{yz} \\
Q^{zx} \\
\end{array}\right)=\left(\begin{array}{c}
+\bm{d}^{\dagger}\lambda_{7}\bm{d}\\
-\bm{d}^{\dagger}\lambda_{5}\bm{d} \\
+\bm{d}^{\dagger}\lambda_{2}\bm{d} \\
-\bm{d}^{\dagger}\lambda_{3}\bm{d} \\
-\bm{d}^{\dagger}\lambda_{8}\bm{d} \\
-\bm{d}^{\dagger}\lambda_{1}\bm{d} \\
-\bm{d}^{\dagger}\lambda_{6}\bm{d} \\
-\bm{d}^{\dagger}\lambda_{4}\bm{d} \\
\end{array}\right).
\end{split}
\end{equation}
Now it is clear that the unitary transformation of the three dimensional complex $\bm{d}$ vector (apart from a global phase factor) gives rise to the underlying $SU(3)$ structure.
Thus the $SU(3)$ algebra of $\{\bm{S},\bm{Q}\}$ can be visualized from Eq.~\eqref{eq:SQSU3d}.
Since the complex $\bm{d}$ vector transfer as a 1-rank tensor under the $SU(3)$ rotations, one can find how $\bm{S}$ and $\bm{Q}$ and other physical quantities 
will transfer under $SU(3)$ as well, which can be written in bilinear or biquadratic terms of $\bm{d}$ and $\bm{\bar{d}}$ in the path integral.

\section{$SU(2)_{\gamma}$ symmetric states/Hamiltonians}
\label{app:SU2state}
In the language of group theory, the three components of $\bm{d}$ belong to the 3-dimensional (3D) fundamental representation $3\equiv(1,0)$ of $SU(3)$ group, and those of $\bm{\bar{d}}$ belong to its complex conjugate representation $\bar{3}\equiv(0,1)$.
So that each $(\bm{\bar{d}},\bm{d})$ bilinear term belongs to the representations $\bar{3}\otimes 3=1\oplus 8$, where $1\equiv(0,0)$ and $8\equiv(1,1)$. 
Explicitly, $|\bm{d}|^2$ belongs to the 1D IR $(0,0)$, and $(\bm{S},\bm{Q})$ belong to the 8D IR $(3,3)$.
Furthermore, each $(\bm{S},\bm{Q})$ bilinear term in Eq.~\eqref{eq:Hgeneric} belongs to the representations $(1,1)\otimes(1,1)=(0,0)\oplus(1,1)\oplus(1,1)\oplus(3,0)\oplus(0,3)\oplus(2,2)$. 
Therefore, we are able to classify the terms in Eq.~\eqref{eq:Hgeneric} according to group theory and find possible spin Hamiltonians respecting the hidden $SU(2)$ symmetries.

Begin with $\bm{d}$ vector and its complex conjugate $\bm{\bar{d}}$, the Cartesian coordinate representation of the three spin-1 states is isomorphic to $SU(3)$ IR $(1,0)$,
\begin{equation}
\renewcommand\arraystretch{1.5}
\begin{array}{lllll}
|(1,0),(1,0)\rangle & \longleftrightarrow & |x\rangle & \longleftrightarrow & d^{x}, \\
|(1,0),(-1,1)\rangle & \longleftrightarrow & |z\rangle & \longleftrightarrow & d^{z}, \\
|(1,0),(0,-1)\rangle & \longleftrightarrow & |y\rangle & \longleftrightarrow & d^{y},
\end{array}
\renewcommand\arraystretch{1}
\end{equation}
and its complex conjugate representation $(0,1)$,
\begin{equation}
\renewcommand\arraystretch{1.5}
\begin{array}{llrrr}
|(0,1),(0,1)\rangle & \longleftrightarrow & \langle y| & \longleftrightarrow & \bar{d}^{y},\\
|(0,1),(1,-1)\rangle & \longleftrightarrow & -\langle z| & \longrightarrow & -\bar{d}^{z},\\
|(0,1),(-1,0)\rangle &\longleftrightarrow & \langle x| & \longleftrightarrow & \bar{d}^{x},
\end{array}
\renewcommand\arraystretch{1}
\end{equation}
where $|(m,n),(p,q)\rangle$ was defined in previous section.
Then $(\bm{S},\bm{Q})$ can be obtained through $(0,1)\otimes(1,0)=(0,0)\oplus(1,1)$, which belong to the 8D IR $(1,1)$,
\begin{equation}\label{eq:AFQ2IR}
\renewcommand\arraystretch{1.5}
\begin{array}{lcc}
|(1,1),(1,1)\rangle & \longleftrightarrow & -\frac{iS^{z}+Q^{xy}}{2},\\
|(1,1),(-1,2)\rangle & \longleftrightarrow & \frac{iS^{x}-Q^{yz}}{2},\\ 
|(1,1),(2,-1)\rangle & \longleftrightarrow & -\frac{iS^{y}-Q^{zx}}{2},\\
|(1,1),(0,0)_{1}\rangle & \longleftrightarrow & \frac{-\sqrt{6}Q^{3z^{2}-r^{2}}-\sqrt{2}Q^{x^{2}-y^{2}}}{4},\\
|(1,1),(0,0)_{2}\rangle & \longleftrightarrow & \frac{-\sqrt{2}Q^{3z^{2}-r^{2}}+\sqrt{6}Q^{x^{2}-y^{2}}}{4},\\
|(1,1),(1,-2)\rangle & \longleftrightarrow & \frac{iS^{x}+Q^{yz}}{2},\\ 
|(1,1),(-2,1)\rangle & \longleftrightarrow & -\frac{iS^{y}+Q^{zx}}{2},\\
|(1,1),(-1,-1)\rangle & \longleftrightarrow & \frac{iS^{z}-Q^{xy}}{2}.
\end{array}
\renewcommand\arraystretch{1}    
\end{equation} 

\begin{figure*}
	\centering
	\includegraphics[scale=0.3]{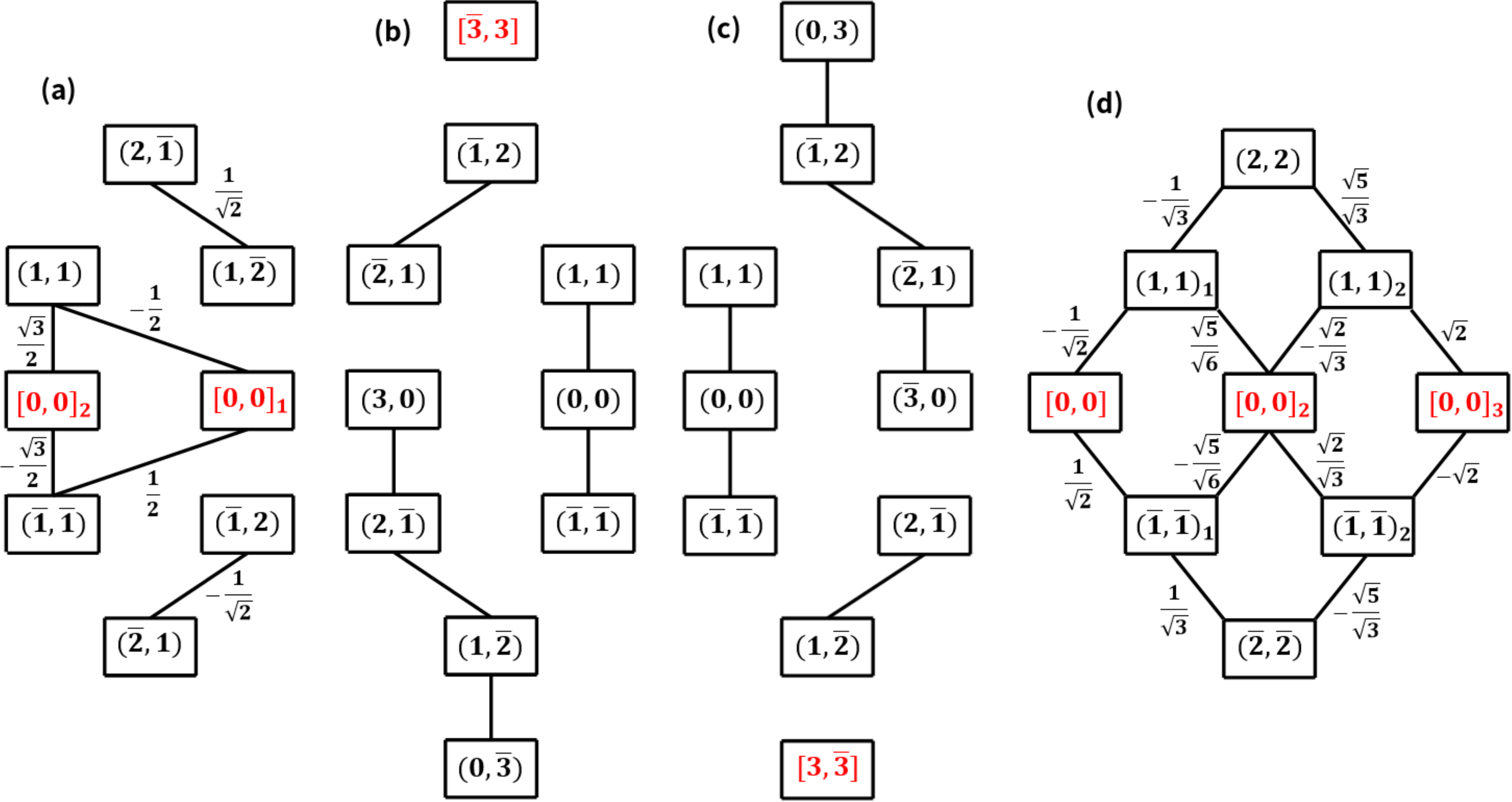}
	\caption{The block diagrams are graphical notation of representations (1,1), (3,0), (0,3) and (2,2). The black solid lines between two block denotes the raising operator $E_{\gamma}$ (upward) or lowering operator $E_{-\gamma}$ (downward). The states marked in red can be utilized to construct $SU(2)_{\gamma}$ symmetric states which will be annihilated by $E_{\gamma}$ and $E_{-\gamma}$.}\label{fig:block}
\end{figure*}

In what follows, we shall construct $SU(2)_{\bm{\gamma}}$ symmetric two-body interactions in terms of bilinear forms of $(\bm{S},\bm{Q})$. 
As mentioned, such bilinear forms belong to the representations $(1,1)\otimes(1,1)=(0,0)\oplus(1,1)\oplus(1,1)\oplus(3,0)\oplus(0,3)\oplus(2,2)$. 
Firstly we shall find out all the $SU(2)_{\bm{\gamma}}$ symmetric states in the IR decomposition of $(1,1)\otimes(1,1)$, which would be annihilated by both the raising operator $E_{\gamma}$ and the lowering operator $E_{-\gamma}$. 
According to the block diagram shown in FIG.~\ref{fig:block}, there exist six linear independent states in the IR decomposition.
We list six linear independent self-conjugate $SU(2)_{\bm{\gamma}}$ symmetric states as follows,
\begin{equation}\label{eq:Gamma}
\renewcommand\arraystretch{1.5}
\begin{array}{lll}
\Gamma_{0}&=&|(0,0),(0,0)\rangle,\\
\Gamma_{1}&=&|(1,1),(0,0)_{2}\rangle_{S}+\sqrt{3}|(1,1),(0,0)_{1}\rangle_{S},\\
\Gamma_{2}&=&|(1,1),(0,0)_{2}\rangle_{A}+\sqrt{3}|(1,1),(0,0)_{1}\rangle_{A},\\
\Gamma_{3}&=&|(3,0),(-3,3)\rangle+|(0,3),(3,-3)\rangle,\\
\Gamma_{4}&=&i|(3,0),(-3,3)\rangle-i|(0,3),(3,-3)\rangle,\\
\Gamma_{5}&=&\sqrt{3}|(2,2),(0,0)_{2}\rangle-|(2,2),(0,0)_{3}\rangle \\
& &-{\sqrt{5}}|(2,2),(0,0)_{1}\rangle.
\end{array}
\renewcommand\arraystretch{1}
\end{equation}
Note that we have already made the bilinear forms symmetrized or antisymmetrized, where $|(1,1),(0,0)_{2}\rangle_{S}$ is a symmetrized state and $|(1,1),(0,0)_{2}\rangle_{A}$ is an antisymmetric state.  
All the possible $SU(2)_{\bm{\gamma}}$ symmetric states can be written as a linear combination of $\Gamma_{n},n=0,\cdots,5$ in Eq.~\eqref{eq:Gamma}. 
Expanding $\Gamma_{n}$ in terms of $(1,1)\otimes(1,1)$ states, $|(1,1),(p_1,q_1)\rangle\otimes|(1,1),(p_2,q_2)\rangle$, through $SU(3)$ Clebsch-Gordan coefficients, 
and replacing abstract states $|(1,1),(p,q)\rangle$ by physical operators $(\bm{S},\bm{Q})$, eventually we obtain all the $SU(2)_{\gamma}$ symmetric spin Hamiltonians.

With the help of $SU(3)$ Clebsch-Gordan coefficients, the $SU(2)_{\gamma}$ symmetric states in Eq.~\eqref{eq:Gamma} can be re-expressed in terms of states in IR $(1,1)$ as follows,
\begin{widetext}
	\begin{equation}\label{eq:Gamma11}
	\begin{split}
	\Gamma_{0}&=\frac{1}{\sqrt{8}}\Big{(}|[0,0]_{1}\rangle_{i}|[0,0]_{1}\rangle_{j}+|[0,0]_{2}\rangle_{i}|[0,0]_{2}\rangle_{j}\Big{)}\\&\quad+\frac{1}{\sqrt{8}}\Big{(}|[1,1]\rangle_{i}|[-1,-1]\rangle_{j}-|[-1,2]_{2}\rangle_{i}|[1,-2]_{2}\rangle_{j}-|[2,-1]_{2}\rangle_{i}|[-2,1]_{2}\rangle_{j}+(i\leftrightarrow{}j)\Big{)}\\
	\Gamma_{1}&=\frac{1}{\sqrt{20}}\Big{(}2|[2,-1]\rangle_{i}|[-2,1]\rangle_{j}+2|[-1,2]\rangle_{i}|[1,-2]\rangle_{j}+4|[1,1]\rangle_{i}|[-1,-1]\rangle_{j}+\left(i\leftrightarrow{}j\right)\Big{)}\\&\quad+\frac{1}{\sqrt{20}}\left(-2|[0,0]_{1}\rangle_{i}|[0,0]_{1}\rangle_{j}+2|[0,0]_{2}\rangle_{i}|[0,0]_{2}\rangle_{j}-2\sqrt{3}|[0,0]_{1}\rangle_{i}|[0,0]_{2}\rangle_{j}-2\sqrt{3}|[0,0]_{2}\rangle_{i}|[0,0]_{1}\rangle_{j}\right),
	\\
	\Gamma_{2}&=\Big{(}|[2,-1]\rangle_{i}|[-2,1]\rangle_{j}-|[-1,2]\rangle_{i}|[1,-2]\rangle_{j}-\left(i\leftrightarrow{}j\right)\Big{)},
	\\
	\Gamma_{3}&=\frac{1}{\sqrt{2}}\Big{(}|[-1,2]\rangle_{i}|[-2,1]\rangle_{j}+|[1,-2]\rangle_{i}|[2,-1]\rangle_{j}-\left(i\leftrightarrow{}j\right)\Big{)},
	\\
	\Gamma_{4}&=\frac{i}{\sqrt{2}}\Big{(}|[-1,2]\rangle_{i}|[-2,1]\rangle_{j}-|[1,-2]\rangle_{i}|[2,-1]\rangle_{j}-\left(i\leftrightarrow{}j\right)\Big{)},\\
	\Gamma_{5}&=\frac{1}{\sqrt{40}}\Big{(}3|[2,-1]\rangle_{i}|[-2,1]\rangle_{j}+3|[-1,2]\rangle_{i}|[1,-2]\rangle_{j}+|[1,1]\rangle_{i}|[-1,-1]\rangle_{j}+\left(i\leftrightarrow{}j\right)\Big{)}\\&\quad+\frac{1}{\sqrt{40}}\left(7|[0,0]_{1}\rangle_{i}|[0,0]_{1}\rangle_{j}+3|[0,0]_{2}\rangle_{i}|[0,0]_{2}\rangle_{j}+2\sqrt{3}|[0,0]_{1}\rangle_{i}|[0,0]_{2}\rangle_{j}+2\sqrt{3}|[0,0]_{2}\rangle_{i}|[0,0]_{1}\rangle_{j}\right),
	\end{split}
	\end{equation}
\end{widetext}
where we use $|[p,q]\rangle_{i}$ to denote state $|(1,1),(p,q)\rangle$ at site $i$ for simplicity. Putting Eq.~\eqref{eq:AFQ2IR} into Eq.~\eqref{eq:Gamma11}, we obtain Eq.~\eqref{eq:H1-6} in the main text. 

\section{Flavor-wave theory}\label{app:FW}

In this appendix, we provide details for the flavor-wave theory \cite{bbqmodel2,flavor1, flavor2,flavor3,flavor4}.
In order to study low energy excitations, we assign three flavors of Shwinger bosons $a_{n\alpha}(j)$ at each site $j$ on the $n^{th}$ sublattice, 
where $\alpha=x,y,z$ corresponds to $x,y,z$ spin states defined in Eq.~\eqref{eq:xyz}. Here $n=1$ for the uniform states, while $n=1,2$ for the bipartite-lattice ordered states.
Thus, the operators $(\bm{S},\bm{Q})$ can be written bilinearly in terms of Schwinger bosons,
\begin{subequations}
	\begin{eqnarray}
	S^{\alpha}_{n}(j)&=&-i\sum_{\beta\gamma}\epsilon_{\alpha\beta\gamma}a^{\dagger}_{n\beta}(j)a_{n\gamma}(j),\\ 
	Q^{\mu\nu}_{n}(j)|_{\mu\neq\nu}&=&-[a^{\dagger}_{n\mu}(j)a_{n\nu}(j)+a^{\dagger}_{n\nu}(j)a_{n\mu}(j)],\\ 
	Q^{x^{2}-y^{2}}_{n}(j)&=&-[a^{\dagger}_{nx}(j)a_{nx}(j)-a^{\dagger}_{ny}(j)a_{ny}(j)], \\ 
	Q^{3z^{2}-r^{2}}_{n}(j)&=&-\frac{1}{\sqrt{3}}[3a^{\dagger}_{nz}(j)a_{nz}(j)-1],
	\end{eqnarray}
\end{subequations}
where the single occupancy constraint
\begin{equation}\label{eq:single-occupancy}
\sum_{\alpha}a_{n\alpha}^{\dagger}(j)a_{n\alpha}(j)=1
\end{equation} 
is imposed.

To obtain various spin ordered states, we shall condense these Schwinger bosons at some components.
Without loss of generality, the condensate components are constructed by an $SU(3)$ rotation $\Omega_n$ in the $n$-th sublattice, which is defined as follows,
\begin{equation}
\left(\begin{array}{c}
a_{n\tilde{x}}\\
a_{n\tilde{y}}\\
a_{n\tilde{z}}\\
\end{array}\right)=\Omega_{n}\left(\begin{array}{c}
a_{nx}\\
a_{ny}\\
a_{nz}\\
\end{array}\right).
\end{equation}
Such an $SU(3)$ rotation $\Omega_{n}$ is site-independent 
and determined by corresponding mean-field $\bm{d}$ vectors and enable us to attribute the condensate to $a_{n\tilde{x}}$ component only.
And $a_{n\tilde{y}}$ and $a_{n\tilde{z}}$ components are thought as small fractions. Then the low energy Hamiltonian can be bilinearized by the Holstein-Primakoff transformation. 
Approximately, $a_{n\tilde{x}}^{\dagger}(j)$ and $a_{n\tilde{x}}(j)$ can be written as,
\begin{equation}
a_{n\tilde{x}}^{\dagger}(j) = a_{n\tilde{x}}(j) = \sqrt{M-a_{n\tilde{y}}^{\dagger}(j)a_{n\tilde{y}}(j)-a_{n\tilde{z}}^{\dagger}(j)a_{n\tilde{z}}(j)}, 
\end{equation}
where $M=1$ in present case considering the single occupancy constraint.

Then we carry out the $1/M$ expansion in the Holstein-Primakoff bosons $a_{n\tilde{y}}$ and $a_{n\tilde{z}}$ up to quadratic order,
and perform the Fourier transformation $a_{n\tilde{\alpha}}(\bm{k}) = \sum_{j}e^{i \bm{k}\cdot \vec{r}_j} a_{n\tilde{\alpha}}(j)/\sqrt{N}$ to obtain the Hamiltonian $\mathcal{H}$ in the $k$-space, 
where $\vec{r}_j$ is the position of the lattice site $j$ and $N$ is the number of magnetic unit cells.
Thus the $k$-space Hamiltonian can be diagonalized by the bosonic Bogoliubov transformation,
\begin{equation}
\mathcal{H}=\sum\limits_{m,\bm{k}}\omega_{m}(\bm{k})b^{\dagger}_{m}(\bm{k})b_{m}(\bm{k})+\mathcal{C},
\end{equation}
where $\omega_{m}(\bm{k})$ is the energy dispersion of $m$-th branch flavor wave, $b_{m}(\bm{k})$ and $b^{\dagger}_{m}(\bm{k})$ 
are bosonic Bogoliubov quasiparticle annihilation and creation operators, and the constant $\mathcal{C}$ does not depend on boson fields.
For uniform states, say, FQ states, $m=1,2$; while for AFQ states, $m=1,2,3,4$. 
As long as the ground states of $\mathcal{H}$ determined by $K_{1}$, $K_{2}$ and $K_{3}$ are given, we are able to obtain the dispersions $\omega_{m}(\bm{k})$ simultaneously.

$\bm{\mbox{FQ1 phase}}\qquad$ For FQ1 phase, the d-vector of ground state reads $\bm{d}^{\tiny{\mbox{FQ1}}}=
\left(\begin{smallmatrix}
1\\
0\\
0
\end{smallmatrix}\right)
$, and the global rotational matrix $\Omega^{\tiny{\mbox{FQ1}}}$ is a $3\times{}3$ unit matrix. We introduce the $SU(3)$ Schwinger bosons as
\begin{equation}
\left(\begin{array}{c}
{a}_{\tilde{x}}\\
{a}_{\tilde{y}}\\
{a}_{\tilde{z}}\\
\end{array}\right)=\left(\begin{array}{c}
\sqrt{n_{x}}\\
a_{y}\\
a_{z}\\
\end{array}\right),
\end{equation}
where
\begin{equation}
\sqrt{n_{x}}\equiv\sqrt{1-a_{y}^{\dagger}a_{y}-a_{z}^{\dagger}a_{z}}
\end{equation}
Expanding spin dipolar and quadrupolar operators $\bm{S}$ and $\bm{Q}$ up to quadratic order of $a_{y}$ and $a_{z}$ gives rise to
\begin{equation}\label{eq:fq1sq}
\begin{split}
&S^{x}=-i(a^{\dagger}_{y}a_{z}-a_{z}^{\dagger}a_{y})\\
&S^{y}=-i(a^{\dagger}_{z}-a_{z})\\
&S^{z}=i(a^{\dagger}_{y}-a_{y})\\
&Q^{xy}=-(a^{\dagger}_{y}+a_{y})\\
&Q^{yz}=-(a^{\dagger}_{y}a_{z}+a_{z}^{\dagger}a_{y})\\
&Q^{zx}=-(a^{\dagger}_{z}+a_{z})\\
&Q^{x^{2}-y^{2}}=-(1-2a^{\dagger}_{y}a_{y}-a^{\dagger}_{z}a_{z})\\
&Q^{3z^{2}-r ^{2}}=\frac{1}{\sqrt{3}}(3a^{\dagger}_{z}a_{z}-1).
\end{split}
\end{equation}
Put them into the Hamiltonian and keep all the terms up to quadratic order of $a_y$ and $a_z$, we obtain
\begin{equation}
\mathcal{H}^{\tiny{\mbox{FQ1}}}=\sum_{\bm{k}}\sum_{m=1}^{2}\omega_{m}^{\tiny{\mbox{FQ1}}}(\bm{k})b_{m}^{\dagger}(\bm{k}){b}_{m}(\bm{k}),
\end{equation}
where
\begin{equation}\label{eq:fq1b}
\begin{split}
&\begin{array}{ll}
b_{1}(\bm{k}) = a_{y}(\bm{k}), & b_{2}(\bm{k}) = a_{z}(\bm{k}),
\end{array}\\
\end{split}
\end{equation} 
Here $\omega^{\tiny{\mbox{\tiny{FQ1}}}}_{1,2}(\bm{k})$ are given in Table \ref{tab:summarysix} in the main text.
Note that all the spins condense at the $|x\rangle$ state. So that $b_{1}^{\dagger}=a_y^{\dagger}$ creates a $|y\rangle$ state and must annihilates an $|x\rangle$ state simultaneously to satisfy the single occupancy constraint in Eq.~\eqref{eq:single-occupancy}.   
It means that the $\omega^{\tiny{\mbox{\tiny{FQ1}}}}_{1}(\bm{k})$ mode corresponds to a two-magnon excitation.
Similarly, $\omega^{\tiny{\mbox{\tiny{FQ1}}}}_{2}(\bm{k})$ causes an $|x\rangle\to|z\rangle$ transition and is a one-magnon mode.

$\bm{\mbox{FQ2 phase}}\qquad$ Considering the d-vector of a FQ2 state being the form of $d^{\tiny{\mbox{FQ2}}}=\left(\begin{smallmatrix}
\cos\vartheta\\
0\\
\sin\vartheta\\
\end{smallmatrix}\right)$ and the global rotational matrix 
\begin{equation}
\Omega^{\tiny{\mbox{FQ2}}}=\left(\begin{array}{ccc}
\cos\vartheta & 0 & -\sin\vartheta\\
0 & 1 & 0 \\
\sin\vartheta & 0 & \cos\vartheta\\
\end{array}\right),
\end{equation}
we introduce rotated Schwinger bosons as follows, 
\begin{equation}
\left(\begin{array}{c}
{a}_{\tilde{x}}\\
{a}_{\tilde{y}}\\
{a}_{\tilde{z}}\\
\end{array}\right)=\left(\begin{array}{c}
\cos\vartheta\sqrt{n_{x}}-\sin\vartheta{}a_{z}\\
a_{y}\\
\sin\vartheta\sqrt{n_{x}}+\cos\vartheta a_{z}
\end{array}\right).
\end{equation}
where $\sin\vartheta$ is determined by the mean-field theory and given in the caption in Table \ref{tab:summarysix} in the main text.
Similiarly, the operators $(\bm{S,Q})$ can be expanded to quadratic order of $a_{{y}}$ and $a_{{z}}$ as follows,  
\begin{equation}\label{eq:fq2sq}
\begin{split}
&S^{x}=-i\sin\vartheta(a^{\dagger}_{y}-a_{y})-i\cos\vartheta(a_{y}^{\dagger}a_{z}-a_{z}^{\dagger}a_{y}),\\
&S^{y}=-i(a^{\dagger}_{z}-a_{z}),\\
&S^{z}=i\cos\vartheta(a^{\dagger}_{y}-a_{y})-i\sin\vartheta(a_{y}^{\dagger}a_{z}-a_{z}^{\dagger}a_{y}),\\
&Q^{xy}=-\cos\vartheta(a^{\dagger}_{y}+a_{y})+\sin\vartheta(a_{y}^{\dagger}a_{z}+a_{z}^{\dagger}a_{y}),\\
&Q^{yz}=-\sin\vartheta(a^{\dagger}_{y}+a_{y})+\sin\vartheta(a_{y}^{\dagger}a_{z}+a_{z}^{\dagger}a_{y}),\\
&Q^{zx}=(\sin^{2}\vartheta-\cos^{2}\vartheta)(a_{z}+a_{z}^{\dagger})\\&\qquad\quad-2\sin\vartheta\cos\vartheta(1-a^{\dagger}_{y}a_{y}-2a^{\dagger}_{z}a_{z}),\\
&Q^{x^{2}-y^{2}}=-\cos^{2}\vartheta+\cos\vartheta\sin\vartheta(a_{z}+a_{z}^{\dagger})\\&\qquad+(1+\cos^{2}\vartheta)a^{\dagger}_{y}a_{y}+(2\cos^{2}\vartheta-1)a_{z}^{\dagger}a_{z},\\
&Q^{3z^{2}-r^{2}}=\sqrt{3}\cos^{2}\vartheta\left(a^{\dagger}_{z}a_{z}-\tan^{2}\vartheta(a^{\dagger}_{y}a_{y}+a^{\dagger}_{z}a_{z})\right)\\&\qquad+\frac{1}{\sqrt{3}}\left(3\sin\vartheta\cos\vartheta(a_{z}+a_{z}^{\dagger})+3\sin^{2}\vartheta-1\right).\\
\end{split}
\end{equation}
Finally we obtain the diagonalized Hamiltonian  
\begin{equation}
\begin{split}
\mathcal{H}^{\tiny{\mbox{FQ2}}}&=\sum_{\bm{k}}\sum_{m=1}^{2}\omega_{m}^{\tiny{\mbox{FQ2}}}(\bm{k})b_{m}^{\dagger}(\bm{k})b_{m}(\bm{k}),
\end{split}
\end{equation}
where $\omega^{\mbox{\tiny{FQ2}}}_{1,2}(\bm{k})$ are given in Table \ref{tab:summarysix} and the Bogoliubov transformation reads 
\begin{equation}
\begin{split}
&a_{y}(\bm{k}) = b_{1}(\bm{k}),\\
&a_{z}(\bm{k})=\cosh(\rho_{\bm{k}}^{\mbox{\tiny{FQ2}}})b_{2}(\bm{k})+\sinh(\rho_{\bm{k}}^{\mbox{\tiny{FQ2}}})b_{2}^{\dagger}(-\bm{k}),
\end{split}
\end{equation} 
with 
\begin{equation}
\exp(\rho_{\bm{k}}^{\mbox{\tiny{FQ2}}})=\sqrt{\frac{1-\gamma(\bm{k})}{1+B_{K}\gamma(\bm{k})}},
\end{equation}
where $B_K$ and $\gamma(\bm{k})$ are given in Table \ref{tab:summarysix}. In this case, condensate components are of $|x\rangle$ and $|z\rangle$ spins. Such that $\omega^{\tiny{\mbox{\tiny{FQ2}}}}_{2}(\bm{k})$ mode corresponds to $|x\rangle\leftrightarrow|z\rangle$ transition, and is a one-magnon mode, while $\omega^{\tiny{\mbox{\tiny{FQ2}}}}_{1}(\bm{k})$ mode corresponds to $|x\rangle+\tan\vartheta{}|z\rangle\leftrightarrow|y\rangle$ transition, and is an admixture of one-magnon and two-magnon modes.

$\bm{\mbox{FQ3 phase}}\qquad$ Now the d-vector is  $\bm{d}^{\mbox{\tiny{FQ3}}}=\left(\begin{smallmatrix}
0\\
0\\
1
\end{smallmatrix}\right)$, and the global rotational matrix reads
\begin{equation}
\Omega^{\tiny{\mbox{FQ3}}}=\left(\begin{array}{ccc}
0 & 0 & -1\\
0 & 1 & 0 \\
1 & 0 & 0\\
\end{array}\right).
\end{equation} 
Then the rotated Schwinger bosons becomes 
\begin{equation}
\left(\begin{array}{c}
{a}_{\tilde{x}}\\
{a}_{\tilde{y}}\\
{a}_{\tilde{z}}\\
\end{array}\right)=\left(\begin{array}{c}
-a_{z}\\
a_{y}\\
\sqrt{n_{x}}\\
\end{array}\right).
\end{equation}
And the operators $\bm{S}$ and $\bm{Q}$ read
\begin{equation}\label{eq:fq3sq}
\begin{array}{l}
S^{x}=-i(a^{\dagger}_{y}-a_{y}),\\
S^{y}=-i(a^{\dagger}_{z}-a_{z}),\\
S^{z}=i(a_{z}^{\dagger}a_{y}-h.c),\\
Q^{xy}=a_{z}^{\dagger}a_{y}+h.c,\\
Q^{yz}=-(a^{\dagger}_{y}+a_{y}),\\
Q^{zx}=a^{\dagger}_{z}+a_{z},\\
Q^{x^{2}-y^{2}}=-(a^{\dagger}_{z}a_{z}-a^{\dagger}_{y}a_{y}),\\
Q^{3z^{2}-r^{2}}_{A}=-\frac{1}{\sqrt{3}}(3a_{y}^{\dagger}a_{y}+3a_{z}^{\dagger}a_{z}-2).\\
\end{array}
\end{equation} 
Put them into the Hamiltonian we obtain
\begin{equation}
\begin{split}
\mathcal{H}^{\tiny{\mbox{FQ3}}}&=\sum_{\bm{k}}\sum_{m=1}^{2}\omega^{\mbox{\tiny{FQ3}}}_{m}(\bm{k})b_{m}^{\dagger}(\bm{k})b_{m}(\bm{k}),
\end{split}
\end{equation}
where $\omega^{\mbox{\tiny{FQ3}}}_{1,2}(\bm{k})$ are given in Table \ref{tab:summarysix} and the Bogoliubov transformation reads
\begin{equation}
\begin{split}
&\begin{array}{cc}
a_{y}(\bm{k})=b_{1}(\bm{k}), & a_{z}(\bm{k})=b_{2}(\bm{k})
\end{array}.
\end{split}
\end{equation}
In this case, the condensate component is $|z\rangle$. Such that $\omega^{\tiny{\mbox{\tiny{FQ3}}}}_{1}(\bm{k})$ mode gives rise to $|z\rangle\to|y\rangle$ transition and is a one-magnon mode, and $\omega^{\tiny{\mbox{\tiny{FQ3}}}}_{2}(\bm{k})$ mode gives rise to $|z\rangle \leftrightarrow |x,y \rangle$ transition and is a one-magnon mode too.

$\bm{\mbox{AFQ1 phase}}\qquad$ The d-vectors in sublattices 1 and 2 are of the form $\bm{d}^{\tiny{\mbox{AFQ1}}}_{1}=
\left(\begin{smallmatrix}
1\\
0\\
0
\end{smallmatrix}\right),\bm{d}^{\tiny{\mbox{AFQ1}}}_{2}=
\left(\begin{smallmatrix}
0\\
1\\
0
\end{smallmatrix}\right)
$, and the corresponding global rotational matrices $\Omega^{\tiny{\mbox{AFQ1}}}_{1}$ and $\Omega^{\tiny{\mbox{AFQ1}}}_{2}$ read,
\begin{equation}
\begin{split}
&\Omega^{\tiny{\mbox{AFQ1}}}_{1}=\left(\begin{array}{ccc}
1 & 0 & 0\\
0 & 1 & 0 \\
0 & 0 & 1\\
\end{array}\right),\quad\Omega^{\tiny{\mbox{AFQ1}}}_{2}=\left(\begin{array}{ccc}
0 & -1 & 0\\
1 & 0 & 0 \\
0 & 0 & 1\\
\end{array}\right).\\
\end{split}
\end{equation}
Therefore the $SU(3)$ Schwinger bosons in the rotated representation can be written as
\begin{equation}
\begin{split}
&\left(\begin{array}{c}
{a}_{1\tilde{x}}\\
{a}_{1\tilde{y}}\\
{a}_{1\tilde{z}}\\
\end{array}\right)=\left(\begin{array}{c}
\sqrt{n_{1x}}\\
a_{1y}\\
a_{1z}\\
\end{array}\right),\\&\left(\begin{array}{c}
{a}_{2\tilde{x}}\\
{a}_{2\tilde{y}}\\
{a}_{2\tilde{z}}\\
\end{array}\right)=\left(\begin{array}{c}
-a_{2y}\\
\sqrt{n_{2x}}\\
a_{2z}\\
\end{array}\right),
\end{split}
\end{equation}
where $\sqrt{n_{mx}}=\sqrt{1-a_{my}^{\dagger}a_{my}-a_{mz}^{\dagger}a_{mz}}$ for $m=1,2$.
Expanding $(\bm{S,Q})$ to quadratic order of $a_{m{y}}$ and $a_{m{z}}$ in each sublattice gives rise to
\begin{equation}\label{eq:afq1sq}
\begin{split}
&\begin{array}{ll}
S^{x}_{1}=-i(a^{\dagger}_{1y}a_{1z}-a_{1z}^{\dagger}a_{1y}), & S^{x}_{2}=i(a^{\dagger}_{2z}-a_{2z}), \\
S^{y}_{1}=-i(a^{\dagger}_{1z}-a_{1z}), & S^{y}_{2}=-i(a^{\dagger}_{2y}a_{2z}-a_{2z}^{\dagger}a_{2y}),\\
S^{z}_{1}=i(a^{\dagger}_{1y}-a_{1y}), & S^{z}_{2}=i(a^{\dagger}_{2y}-a_{2y}),\\
Q^{xy}_{1}=-(a^{\dagger}_{1y}+a_{1y}), & Q^{xy}_{2}=(a^{\dagger}_{2y}+a_{2y}),  \\
Q^{yz}_{1}=-(a^{\dagger}_{1y}a_{1z}+a_{1z}^{\dagger}a_{1y}), & Q^{yz}_{2}=-(a^{\dagger}_{2z}+a_{2z}), \\
Q^{zx}_{1}=-(a^{\dagger}_{1z}+a_{1z}), & Q^{zx}_{2}=(a^{\dagger}_{2y}a_{2z}+2_{2z}^{\dagger}2_{2y}),\\
\end{array}\\
&\begin{array}{l}
Q^{x^{2}-y^{2}}_{1}=-(1-2a^{\dagger}_{1y}a_{1y}-a^{\dagger}_{1z}a_{1z}), \\ Q^{x^{2}-y^{2}}_{2}=(1-2a^{\dagger}_{2y}a_{2y}-a^{\dagger}_{2z}a_{2z}),\\
Q^{3z^{2}-r^{2}}_{1}=\frac{1}{\sqrt{3}}(3a^{\dagger}_{1z}a_{1z}-1), \\	 Q^{3z^{2}-r^{2}}_{2}=\frac{1}{\sqrt{3}}(3a^{\dagger}_{2z}a_{2z}-1). \\
\end{array}
\end{split}
\end{equation}
Then the mean-field Hamiltonian of AFQ1 becomes
\begin{equation}
\mathcal{H}^{\tiny{\mbox{AFQ1}}}=\sum_{\bm{k}}\sum_{m=1}^{4}\omega_{m}^{\tiny{\mbox{AFQ1}}}(\bm{k})b_{m}^{\dagger}(\bm{k})b_{m}(\bm{k}),
\end{equation}
where $\omega^{\tiny{\mbox{\tiny{FQ1}}}}_{1,2}(\bm{k})$ are given in Table \ref{tab:summarysix} in the main text. The Bogoliubov transformation reads
\begin{equation}
\begin{split}
&a_{1z}(\bm{k})=b_{3}(\bm{k}),\quad{}a_{2z}(\bm{k})=b_{4}(\bm{k}),\\
&\left(\begin{array}{c}
a_{1y}(\bm{k})\\
a_{2y}(\bm{k})\\
\end{array}\right)=\left(\begin{array}{llll}
c_{\bm{k}}^{\tiny{\mbox{A1}}} & c_{\bm{k}}^{\tiny{\mbox{A1}}} & s_{\bm{k}}^{\tiny{\mbox{A1}}} & -s_{\bm{k}}^{\tiny{\mbox{A1}}} \\
c_{\bm{k}}^{\tiny{\mbox{A1}}} & -c_{\bm{k}}^{\tiny{\mbox{A1}}} & s_{\bm{k}}^{\tiny{\mbox{A1}}} & s_{\bm{k}}^{\tiny{\mbox{A1}}}\\
\end{array}\right)\left(\begin{array}{l}
b_{1}(\bm{k})\\
b_{2}(\bm{k}) \\
b_{1}^{\dagger}(-\bm{k})\\
b_{2}^{\dagger}(-\bm{k}) \\
\end{array}\right),
\end{split}
\end{equation}
with
\begin{equation}
c_{\bm{k}}^{\tiny{\mbox{A1}}}\equiv\frac{1}{\sqrt{2}}\cosh(\rho_{\bm{k}}^{\tiny{\mbox{A1}}}),\quad{}s_{\bm{k}}^{\tiny{\mbox{A1}}}\equiv\frac{1}{\sqrt{2}}\sinh(\rho_{\bm{k}}^{\tiny{\mbox{A1}}}),
\end{equation} 
and $\rho_{\bm{k}}^{\tiny{\mbox{A1}}}$ is given as 
\begin{equation}
\exp(2\rho_{\bm{k}}^{\tiny{\mbox{A1}}})=\sqrt{\frac{1+\gamma(\bm{k})}{1-\gamma(\bm{k})}}.
\end{equation}
Similar to the case of FQ1, all the spins condense at the $|x\rangle$ state. So that $\omega^{\tiny{\mbox{\tiny{AFQ1}}}}_{1,2}(\bm{k})$ modes give rise to $|x\rangle\to|y\rangle$ transitions and correspond to two-magnon excitations. And $\omega^{\tiny{\mbox{\tiny{AFQ1}}}}_{3,4}(\bm{k})$ modes give rise to $|x\rangle\leftrightarrow|z\rangle$ transition and are one-magnon excitations.

$\bm{\mbox{AFQ2 phase}}\qquad$ In this phase, the d-vectors in two sublattice are $d^{\tiny{\mbox{AFQ2}}}_{1}=\left(\begin{smallmatrix}
\cos\vartheta\\
0\\
\sin\vartheta\\
\end{smallmatrix}\right)$ and $d^{\tiny{\mbox{AFQ2}}}_{2}=\left(\begin{smallmatrix}
\cos\vartheta\\
0\\
-\sin\vartheta\\
\end{smallmatrix}\right)$, and the global rotational matrices read
\begin{equation}
\begin{split}
&\Omega^{\tiny{\mbox{AFQ2}}}_{1}=\left(\begin{array}{ccc}
\cos\vartheta & 0 & -\sin\vartheta\\
0 & 1 & 0 \\
\sin\vartheta & 0 & \cos\vartheta\\
\end{array}\right),\\
&\Omega^{\tiny{\mbox{AFQ2}}}_{1}=\left(\begin{array}{ccc}
\cos\vartheta & 0 & \sin\vartheta\\
0 & 1 & 0 \\
-\sin\vartheta & 0 & \cos\vartheta\\
\end{array}\right).
\end{split}
\end{equation} 
We have $SU(3)$ Schwinger bosons in such rotated representation as follows, 
\begin{equation}
\begin{split}
&\left(\begin{array}{c}
a_{1\tilde{x}}\\
a_{1\tilde{y}}\\
a_{1\tilde{z}}\\
\end{array}\right)=\left(\begin{array}{c}
\cos\vartheta\sqrt{n_{1x}}-\sin\vartheta{}a_{1z}\\
a_{1y}\\
\sin\vartheta{}\sqrt{n_{1x}}+\cos\vartheta{}a_{1z}
\end{array}\right),\\&\left(\begin{array}{c}
a_{2\tilde{x}}\\
a_{2\tilde{y}}\\
a_{2\tilde{z}}\\
\end{array}\right)=\left(\begin{array}{c}
\cos\vartheta{}\sqrt{n_{2x}}+\sin\vartheta{}a_{2z}\\
a_{2y}\\
-\sin\vartheta{}\sqrt{n_{2x}}+\cos\vartheta{}a_{2z}
\end{array}\right).
\end{split}
\end{equation}
Then the forms of $(\bm{S,Q})$ for each sublattice of AFQ2 are very similar to Eq.~\eqref{eq:fq2sq}, and here we do not list them explicitly.

The corresponding Hamiltonian becomes 
\begin{equation}
\begin{split}
\mathcal{H}^{\tiny{\mbox{AFQ2}}}&=\sum_{\bm{k}}\sum_{m=1}^{4}\omega^{\tiny{\mbox{\tiny{AFQ2}}}}_{m}(\bm{k})b_{m}^{\dagger}(\bm{k})b_{m}(\bm{k}),
\end{split}
\end{equation}
where $\omega_{1,2,3,4}^{\tiny{\mbox{\tiny{AFQ2}}}}$ are given in Table \ref{tab:summarysix} in the main text.
The Bogoliubov transformation are chosen as 
\begin{equation}
\left(\begin{array}{c}
a_{1y}(\bm{k})\\
a_{2y}(\bm{k})\\
\end{array}\right)=\frac{1}{\sqrt{2}}\left(\begin{array}{cc}
1 & 1 \\
1 & -1\\
\end{array}\right)\left(\begin{array}{c}
b_{1}(\bm{k})\\
b_{2}(\bm{k}) \\
\end{array}\right),
\end{equation}
and
\begin{equation}
\left(\begin{array}{c}
a_{1z}(\bm{k})\\
a_{2z}(\bm{k})\\
\end{array}\right)=\left(\begin{array}{cccc}
c_{1\bm{k}}^{\tiny{\mbox{A2}}} & c_{2\bm{k}}^{\tiny{\mbox{A2}}} & s_{1\bm{k}}^{\tiny{\mbox{A2}}} & s_{2\bm{k}}^{\tiny{\mbox{A2}}} \\
c_{1\bm{k}}^{\tiny{\mbox{A2}}} & -c_{2\bm{k}}^{\tiny{\mbox{A2}}} & s_{1\bm{k}}^{\tiny{\mbox{A2}}} & -s_{2\bm{k}}^{\tiny{\mbox{A2}}}\\
\end{array}\right)\left(\begin{array}{c}
b_{3}(\bm{k})\\
b_{4}(\bm{k}) \\
b_{3}^{\dagger}(-\bm{k})\\
b_{4}^{\dagger}(-\bm{k})\\
\end{array}\right),
\end{equation}
where 
\begin{equation}
c_{m\bm{k}}^{\tiny{\mbox{A2}}}\equiv\frac{1}{\sqrt{2}}\cosh(\rho_{{m\bm{k}}}^{\tiny{\mbox{A2}}}),\quad{}s_{m\bm{k}}^{\tiny{\mbox{A2}}}\equiv\frac{1}{\sqrt{2}}\sinh(\rho_{m\bm{k}}^{\tiny{\mbox{A2}}}),
\end{equation} 
with $m=1,2$ and 
\begin{equation}
\begin{split}
&\exp(2\rho_{1\bm{k}}^{\tiny{\mbox{A2}}})=\sqrt{\frac{1+\gamma(\bm{k})}{1-B_{K}\gamma(\bm{k})}},\\
&\exp(2\rho_{2\bm{k}}^{\tiny{\mbox{A2}}})=\sqrt{\frac{1-\gamma(\bm{k})}{1+B_{K}\gamma(\bm{k})}}.
\end{split}
\end{equation}
Similar to the case of FQ2, in this case condensate components are of $|x\rangle$ and $|z\rangle$ spins. So $\omega^{\tiny{\mbox{\tiny{AFQ2}}}}_{3,4}(\bm{k})$ modes which correspond to $|x\rangle\leftrightarrow|z\rangle$ transition are one-magnon modes, while $\omega^{\tiny{\mbox{\tiny{AFQ2}}}}_{1,2}(\bm{k})$ modes correspond to $|x\rangle\pm\tan\vartheta{}|z\rangle\leftrightarrow|y\rangle$ transitions, and are admixtures of one-magnon and two-magnon modes.

$\bm{\mbox{AFQ3 phase}}$  The AFQ3 ground states are given by the d-vectors in two sublattices as the following, $\bm{d}_{1}^{\tiny{\mbox{AFQ3}}}=
\left(\begin{smallmatrix}
0\\
0\\
1
\end{smallmatrix}\right),\bm{d}_{2}^{\tiny{\mbox{AFQ3}}}=
\left(\begin{smallmatrix}
1\\
0\\
0\\
\end{smallmatrix}\right)
$. Now the global rotational matrices $\Omega^{\tiny{\mbox{AFQ3}}}_{1}$ and $\Omega^{\tiny{\mbox{AFQ3}}}_{2}$ read,
\begin{equation}
\begin{split}
&\Omega^{\tiny{\mbox{AFQ3}}}_{1}=\left(\begin{array}{ccc}
0 & 0 & -1\\
0 & 1 & 0 \\
1 & 0 & 0\\
\end{array}\right),\quad\Omega^{\tiny{\mbox{AFQ3}}}_{2}=\left(\begin{array}{ccc}
1 & 0 & 0\\
0 & 1 & 0 \\
0 & 0 & 1\\
\end{array}\right).\\
\end{split}
\end{equation}
The corresponding Schwinger bosons in the rotated representations are
\begin{equation}
\begin{split}
&\left(\begin{array}{c}
a_{1\tilde{x}}\\
a_{2\tilde{y}}\\
a_{3\tilde{z}}\\
\end{array}\right)=\left(\begin{array}{c}
-a_{1z}\\
a_{1y}\\
\sqrt{n_{1x}}\\
\end{array}\right),\\
&\left(\begin{array}{c}
a_{2\tilde{x}}\\
a_{2\tilde{y}}\\
a_{2\tilde{z}}\\
\end{array}\right)=\left(\begin{array}{c}
\sqrt{n_{2x}}\\
a_{2y}\\
a_{2z}\\
\end{array}\right).
\end{split}
\end{equation}
And operators $(\bm{S},\bm{Q})$ for each sublattice read
\begin{equation}\label{eq:afq3sq}
\begin{split}
&\begin{array}{ll}
S^{x}_{1}=-i(a^{\dagger}_{1y}-a_{1y}), & S^{x}_{2}=-i(a^{\dagger}_{2y}a_{2z}- a^{\dagger}_{2z}a_{2y}),\\
S^{y}_{1}=-i(a^{\dagger}_{1z}-a_{1z}), & S^{y}_{2}=-i(a^{\dagger}_{2z}-a_{2z}),\\
S^{z}_{1}=i(a_{1z}^{\dagger}a_{1y}-a_{1y}^{\dagger}a_{1z}), & S^{z}_{2}=i(a^{\dagger}_{2y}-a_{2y}),\\
Q^{xy}_{1}=a_{1z}^{\dagger}a_{1y}+a_{1y}^{\dagger}a_{1z}, & Q^{xy}_{2}=-(a^{\dagger}_{2y}+a_{2y}),  \\
Q^{yz}_{1}=-(a^{\dagger}_{1y}+a_{1y}), & Q^{yz}_{2}=-(a^{\dagger}_{2y}a_{2z}+ a^{\dagger}_{2z}a_{2y}),\\
Q^{zx}_{1}=a^{\dagger}_{1z}+a_{1z}, & Q^{zx}_{2}=-(a^{\dagger}_{2z}+a_{2z}),\\
\end{array}\\
&\begin{array}{l}
Q^{x^{2}-y^{2}}_{1}=-(a^{\dagger}_{1z}a_{1z}-a^{\dagger}_{1y}a_{1y}),\\ Q^{x^{2}-y^{2}}_{2}=-(1-2a^{\dagger}_{2y}a_{2y}-a^{\dagger}_{2z}a_{2z}),\\
Q^{3z^{2}-r^{2}}_{1}=-\frac{1}{\sqrt{3}}(3a_{1z}^{\dagger}a_{1z}+3a_{1y}^{\dagger}a_{1y}-2), \\
Q^{3z^{2}-r^{2}}_{2}=-\frac{1}{\sqrt{3}}(1-3a_{2z}^{\dagger}a_{2z}).\\
\end{array}
\end{split}
\end{equation}
The diagonolized Hamiltonian reads
\begin{equation}
\mathcal{H}^{\tiny{\mbox{AFQ3}}}=\sum_{\bm{k}}\sum_{m=1}^{4}\omega^{\tiny{\mbox{AFQ3}}}_{m}(\bm{k})b_{m}^{\dagger}(\bm{k})b_{m}(\bm{k}).
\end{equation}
where $\omega_{1,2,3,4}^{\tiny{\mbox{\tiny{AFQ3}}}}$ are given in Table \ref{tab:summarysix} and
\begin{equation}
\begin{split}
&a_{1y}(\bm{k})=b_{1}(\bm{k}),\qquad{}a_{2y}(\bm{k})=b_{2}(\bm{k}),\\	
&\left(\begin{array}{l}
a_{1z}(\bm{k})\\
a_{2z}^{\dagger}(-\bm{k})\\
\end{array}\right)=\left(\begin{array}{cc}
\cosh(\rho^{\tiny{\mbox{A3}}}_{\bm{k}}) & \sinh(\rho^{\tiny{\mbox{A3}}}_{\bm{k}})  \\
\sinh(\rho^{\tiny{\mbox{A3}}}_{\bm{k}}) & \cosh(\rho^{\tiny{\mbox{A3}}}_{\bm{k}})  \\
\end{array}\right)\left(\begin{array}{c}
b_{3}(\bm{k})\\
b_{4}^{\dagger}(-\bm{k}) \\
\end{array}\right).
\end{split}
\end{equation}
Here
\begin{equation}
\exp(2\rho^{\tiny{\mbox{A3}}}_{\bm{k}})=\sqrt{\frac{3K_{3}+K_{1}-4K_{2}\gamma(\bm{k})}{3K_{3}+K_{1}+4K_{2}\gamma(\bm{k})}}.
\end{equation} 
In this case all spins  condense at the $|x\rangle$ state. Thus $\omega^{\tiny{\mbox{\tiny{AFQ3}}}}_{1,2}(\bm{k})$ modes corresponding to $|x\rangle\leftrightarrow|y\rangle$ transitions are two-magnon modes. While $\omega^{\tiny{\mbox{\tiny{AFQ3}}}}_{3,4}(\bm{k})$ modes corresponding to $|x\rangle\leftrightarrow|z\rangle$ transitions are one-magnon modes.

\section{Spectral functions}\label{app:sf}
We provide details for spin spectral function $S(\bm{q},\omega)$ and spin quardrupole spectral function $Q(\bm{q},\omega)$, which are calculated by the linearized flavor-wave theory. 

\subsection{Spin spectral functions}\label{app:ssf}

In this subsection, we demonstrate details for $S(\bm{q},\omega)$.

$\bm{\mbox{FQ1 phase}}\quad$ The spin operators in the flavor-wave theory read
\begin{equation}\label{eq:fq1s}
\begin{split}
&S^{x}=(r_{1}-ir_{2})a_{z}+(r_{1}+ir_{2})a^{\dagger}_{z},\\
&S^{y}=(-r_{3}-ir_{0})a_{z}+(-r_{3}+ir_{0})a^{\dagger}_{z},\\
&S^{z}=ua_{y} + u^{*}a^{\dagger}_{y},
\end{split}
\end{equation}
where 
\begin{equation}\label{eq:u}
u=i(r_{0}-ir_{3})^{2}+i(r_{2}+ir_{1})^{2},
\end{equation}
and the quadratic boson and constant terms are omitted. Note that the constant does not contribute to any excitations thereby the spectral functions. Then spin spectral function reads
\begin{equation}
\begin{split}
S_{\mbox{\tiny{FQ1}}}(\bm{q},\omega)=&2\pi\left[\delta(\omega-\omega^{\mbox{\tiny{FQ1}}}_{2}(\bm{q}))+|u|^{2}\delta\left(\omega-\omega^{\mbox{\tiny{FQ1}}}_{1}(\bm{q})\right)\right].\\
\end{split}
\end{equation}

$\bm{\mbox{AFQ1 phase}}\quad$ The spin operators for sublattice 1 are the same as Eq.~\eqref{eq:fq1s} but with additional sublattice subindex and the spin operators for sublattice 2 read
\begin{equation}
\begin{split}
&S^{x}_{2}=(ir_{0}-r_{3})a_{2z}+(-ir_{0}-r_{3})a^{\dagger}_{2z},\\
&S^{y}_{2}=(-r_{1}-ir_{2})a_{2z}+(-r_{1}+ir_{2})a^{\dagger}_{2z},\\
&S^{z}_{2}=-u^{*}a_{2y} - ua^{\dagger}_{2y}.
\end{split}
\end{equation}
And spin spectral function is
\begin{equation}
\begin{split}
S_{\mbox{\tiny{AFQ1}}}(\bm{q},\omega)=&2\pi\delta(\omega-\omega^{\mbox{\tiny{AFQ1}}}_{3}(\bm{q}))\\
+2\pi&\sqrt{\frac{1-\gamma(\bm{q})}{1+\gamma(\bm{q})}}|u|^{2}\delta(\omega-\omega^{\mbox{\tiny{AFQ1}}}_{1}(\bm{q})),\\
\end{split}
\end{equation}
where $u$ is defined in Eq.~\eqref{eq:u}.

$\bm{\mbox{FQ2 phase}}\quad$ The spin operators read
\begin{equation}\label{eq:fq2s}
\begin{split}
&S^{x}=(-ir_{2}+r_{1}\cos{}2\vartheta)a_{z}-\sin\vartheta(r_{3}+ir_{0})a_{y}+h.c.,\\
&S^{y}=(-ir_{0}-{}r_{3}\cos{}2\vartheta)a_{z}-\sin\vartheta(r_{1}-ir_{2})a_{y}+h.c.,\\
&S^{z}=u\cos\vartheta{}a_{y} + (r_{0}r_{1}+r_{2}r_{3})\sin{}2\vartheta{}a_{z}+h.c..\\
\end{split}
\end{equation}
And spin spectral function reads
\begin{equation}
\begin{split}
&S_{\mbox{\tiny{FQ2}}}(\bm{q},\omega)=2\pi{}F_{1}(\vartheta,\hat{r})\delta(\omega-\omega_{1}^{\mbox{\tiny{FQ2}}}(\bm{q}))\\&\qquad+2\pi{}F_{2}(\vartheta,\hat{r})\sqrt{\frac{1+B_{K}\gamma(\bm{q})}{1-\gamma(\bm{q})}}\delta(\omega-\omega_{2}^{\mbox{\tiny{FQ2}}}(\bm{q})),\\
\end{split}
\end{equation}
where $u$ is defined in Eq.~\eqref{eq:u} and
\begin{equation}\label{eq:F1F2}
\begin{split}
&F_{1}(\vartheta,\hat{r})=1+(|u|^{2}-1)\cos^{2}\vartheta,\\
&F_{2}(\vartheta,\hat{r})=1+\frac{4r_{0}^{2}+4r_{2}^{2}-3-|u|^{2}}{4}\sin^{2}2\vartheta.
\end{split}
\end{equation} 

$\bm{\mbox{AFQ2 phase}}\quad$ The forms of spin operators for sublattice 1 are the same as Eq.~\eqref{eq:fq2s}. And for sublattice 2 we can obtain spin operators by taking $\vartheta\rightarrow{}-\vartheta$. So here we do not list them explicitly.  
The dipolar spin spectral function reads
\begin{equation}
\begin{split}
&S_{\mbox{\tiny{AFQ2}}}(\bm{q},\omega)=2\pi\sin^{2}\vartheta\delta(\omega-\omega^{\mbox{\tiny{AFQ2}}}_{2}(\bm{q}))\\
&+2\pi|u|^{2}\cos^{2}\vartheta\delta(\omega-\omega^{\mbox{\tiny{AFQ2}}}_{1}(\bm{q}))
\\
&+2\pi(r_{0}^{2}+r_{2}^{2})\sqrt{\frac{1-B_{K}\gamma(\bm{q})}{1+\gamma(\bm{q})}}\delta(\omega-\omega_{3}^{\mbox{\tiny{AFQ2}}}(\bm{q}))\\
&+2\pi(r_{1}^{2}+r_{3}^{2})\cos^{2}2\vartheta\sqrt{\frac{1+\gamma(\bm{q})}{1-B_{K}\gamma(\bm{q})}}\delta(\omega-\omega_{3}^{\mbox{\tiny{AFQ2}}}(\bm{q})),\\
&+\pi{}\frac{(1-|u|^{2})\sin^{2}2\vartheta}{2}\sqrt{\frac{1-\gamma(\bm{q})}{1+B_{K}\gamma(\bm{q})}}\delta(\omega-\omega_{4}^{\mbox{\tiny{FQ2}}}(\bm{q})),
\end{split}
\end{equation}
where $u$ is defined in Eq.~\eqref{eq:u}.

$\bm{\mbox{FQ3\ phase}}\quad$ The spin operators are 
\begin{equation}\label{eq:fq3s}
\begin{split}
&S^{x}=(ir_{0}+r_{3})a_{y}+(r_{1}+ir_{2})a_{z}+h.c.,\\
&S^{y}=(ir_{0}-{}r_{3})a_{z}+(r_{1}-ir_{2})a_{y}+h.c.,\\
&S^{z}=0.
\end{split}
\end{equation}
And the dipolar spin spectral function reads
\begin{equation}\label{eq:ssffq3}
\begin{split}
&S_{\mbox{\tiny{FQ3}}}(\bm{q},\omega)=2\pi\left[\delta(\omega-\omega^{\mbox{\tiny{FQ3}}}_{1}(\bm{q}))+\delta(\omega-\omega^{\mbox{\tiny{FQ3}}}_{2}(\bm{q}))\right],\\
\end{split}
\end{equation}
Note that $S_{\mbox{\tiny{FQ3}}}(\bm{q},\omega)$ does not depend on the $\bm{d}$ vector.

$\bm{\mbox{AFQ3\ phase}}\quad$ The forms of spin operators for sublattice 1(2) are the same as Eq.~\eqref{eq:fq3s}(Eq.~\eqref{eq:fq1s}) with additional sublattice index. Thus 
the dipolar spin spectral function for an AFQ3 state does not depend on the $\bm{d}$ vector as well and reads
\begin{equation}\label{eq:ssfafq3}
\begin{split}
&S_{\mbox{\tiny{AFQ3}}}(\bm{q},\omega)=\pi\delta(\omega-\omega^{\mbox{\tiny{AFQ3}}}_{1}(\bm{q}))\\&\qquad\qquad+\pi\sqrt{\frac{C_{K}-D_{K}\gamma(\bm{q})}{C_{K}+D_{K}\gamma(\bm{q})}}\delta(\omega-\omega^{\mbox{\tiny{AFQ3}}}_{3}(\bm{q}))\\&\qquad\qquad+\pi\sqrt{\frac{C_{K}-D_{K}\gamma(\bm{q})}{C_{K}+D_{K}\gamma(\bm{q})}}\delta(\omega-\omega^{\mbox{\tiny{AFQ3}}}_{4}(\bm{q})).\\
\end{split}
\end{equation}

\subsection{Quadrupole spectral functions}\label{app:qsf}

In this subsection, we demonstrate details for $Q(\bm{q},\omega)$.

$\bm{\mbox{FQ1 phase}}\quad$ The $\bm{Q}$ operators read
\begin{equation}\label{eq:fq1q}
\begin{split}
&Q^{xy}=va_{z}+h.c.,\quad  Q^{yz}=(r_{2}+ir_{1})a_{z}+h.c.,\\
&Q^{zx}=(ir_{3}-r_{0})a_{z}+h.c.,\quad   Q^{3z^{2}-r^{2}} = 0,\\
&Q^{x^{2}-y^{2}} = 2(ir_{3}-r_{0})(r_{2}+ir_{1})a_{y}+h.c.,
\end{split}
\end{equation}
where
\begin{equation}\label{eq:v}
v = (ir_{0}+r_{2})^{2} - (ir_{3}-r_{0})^{2}.
\end{equation} 
Notice the $r_{i}$s are defined in Eq.~\eqref{eq:R_ri} and again the quadratic boson and constant terms are omitted.
Then the quadrupolar spin spectral function reads
\begin{equation}
\begin{split}
Q_{\mbox{\tiny{FQ1}}}(\bm{q},\omega)=&2\pi\delta(\omega-\omega^{\mbox{\tiny{FQ1}}}_{2}(\bm{q}))\\&+2\pi(2-|u|^{2})\delta\left(\omega-\omega^{\mbox{\tiny{FQ1}}}_{1}(\bm{q})\right),
\end{split}
\end{equation}
where $u$ is defined in Eq.~\eqref{eq:u}.

$\bm{\mbox{AFQ1 phase}}\quad$ The $\bm{Q}$ operators for sublattice 1 are the same as Eq.~\eqref{eq:fq1q} but with additional sublattice subindex and the $\bm{Q}$  operators for sublattice 2 read
\begin{equation}
\begin{split}
&Q^{xy}_{2}=-v^{*}a_{2z}+h.c.,\quad  Q^{yz}_{2}=-(r_{0}+ir_{3})a_{2z}+h.c.,\\
&Q^{zx}_{2}=(ir_{1}-r_{2})a_{2z}+h.c.,\quad   Q^{3z^{2}-r^{2}}_{2} = 0,\\
&Q^{x^{2}-y^{2}}_{2} = 2(ir_{3}+r_{0})(r_{2}-ir_{1})a_{2y}+h.c.,
\end{split}
\end{equation}
where $v$ is defined in Eq.~\eqref{eq:v}.
The quadrupolar spin spectral function is
\begin{equation}
\begin{split}
Q_{\mbox{\tiny{AFQ1}}}(\bm{q},\omega)=&2\pi\delta(\omega-\omega^{\mbox{\tiny{AFQ1}}}_{3}(\bm{q}))\\
+2\pi(2&-|u|^{2})\sqrt{\frac{1-\gamma(\bm{q})}{1+\gamma(\bm{q})}}\delta(\omega-\omega^{\mbox{\tiny{AFQ1}}}_{1}(\bm{q})),\\
\end{split}
\end{equation}
where $u$ is defined in Eq.~\eqref{eq:u}.

$\bm{\mbox{FQ2 phase}}\quad$ The $\bm{Q}$ operators read
\begin{equation}\label{eq:fq2q}
\begin{split}
&Q^{xy}=v\cos\vartheta{}a_{y}-(r_{0}r_{2}-r_{1}r_{3})\sin{}2\vartheta{}a_{z}+h.c.,\\
&Q^{zx}=(ir_{3}-{}r_{0}\cos{}2\vartheta)a_{z}-\sin\vartheta(ir_{1}+r_{2})a_{y}+h.c.,\\
&Q^{yz}=(ir_{1}+{}r_{2}\cos{}2\vartheta)a_{z}+\sin\vartheta(ir_{3}-r_{0})a_{y}+h.c.,\\
&Q^{x^{2}-y^{2}}=(1/2-r_{1}^{2}-{}r^{2}_{2})\sin{}2\vartheta{}a_{z}\\&\qquad+2\cos\vartheta(ir_{1}+r_{2})(ir_{3}-r_{0})a_{y}+h.c.,\\
&Q^{3z^{2}-r^{2}}=\sqrt{3}\sin{}2\vartheta{}a_{z}/2+h.c..\\
\end{split}
\end{equation}
where $v$ is defined in Eq.~\eqref{eq:v}.
And quadrupolar spin spectral function reads
\begin{equation}
\begin{split}
&Q_{\mbox{\tiny{FQ2}}}(\bm{q},\omega)=2\pi{}F_{3}(\vartheta,\hat{r})\delta(\omega-\omega_{1}^{\mbox{\tiny{FQ2}}}(\bm{q}))\\&\qquad+2\pi{}F_{4}(\vartheta,\hat{r})\sqrt{\frac{1+B_{K}\gamma(\bm{q})}{1-\gamma(\bm{q})}}\delta(\omega-\omega_{2}^{\mbox{\tiny{FQ2}}}(\bm{q})),\\
\end{split}
\end{equation}
where $u$ is defined in Eq.~\eqref{eq:u} and
\begin{equation}\label{eq:F3F4}
\begin{split}
&F_{3}(\vartheta,\hat{r})=1+(1-|u|^{2})\cos^{2}\vartheta,\\
&F_{4}(\vartheta,\hat{r})=1+\frac{4r_{1}^{2}+4r_{3}^{2}-1+|u|^{2}}{4}\sin^{2}2\vartheta.
\end{split}
\end{equation} 

$\bm{\mbox{AFQ2 phase}}\quad$ The forms of $\bm{Q}$ operators for sublattice 1 are the same as Eq.~\eqref{eq:fq2q}. And for sublattice 2 we can obtain the $\bm{Q}$ operators by taking $\vartheta\rightarrow{}-\vartheta$. 
The quadrupolar spin spectral function reads
\begin{equation}
\begin{split}
&Q_{\mbox{\tiny{AFQ2}}}(\bm{q},\omega)=2\pi\sin^{2}\vartheta\delta(\omega-\omega^{\mbox{\tiny{AFQ2}}}_{2}(\bm{q}))\\
&+2\pi(2-|u|^{2})\cos^{2}\vartheta\delta(\omega-\omega^{\mbox{\tiny{AFQ2}}}_{1}(\bm{q}))
\\
&+2\pi(r_{1}^{2}+r_{3}^{2})\sqrt{\frac{1-B_{K}\gamma(\bm{q})}{1+\gamma(\bm{q})}}\delta(\omega-\omega_{3}^{\mbox{\tiny{AFQ2}}}(\bm{q}))\\
&+2\pi(r_{0}^{2}+r_{2}^{2})\cos^{2}2\vartheta\sqrt{\frac{1+\gamma(\bm{q})}{1-B_{K}\gamma(\bm{q})}}\delta(\omega-\omega_{3}^{\mbox{\tiny{AFQ2}}}(\bm{q})),\\
&+\pi{}\frac{(3+|u|^{2})\sin^{2}2\vartheta}{2}\sqrt{\frac{1-\gamma(\bm{q})}{1+B_{K}\gamma(\bm{q})}}\delta(\omega-\omega_{4}^{\mbox{\tiny{FQ2}}}(\bm{q})),
\end{split}
\end{equation}
where $u$ is defined in Eq.~\eqref{eq:u}.

$\bm{\mbox{FQ3\ phase}}\quad$ The $\bm{Q}$ operators are 
\begin{equation}\label{eq:fq3q}
\begin{split}
&Q^{xy}=Q^{x^{2}-y^{2}}=Q^{3z^{2}-r^{2}}=0,\\
&Q^{yz}=(ir_{1}-r_{2})a_{z}+(ir_{3}-{}r_{0})a_{y}+h.c.,\\
&Q^{zx}=(ir_{3}+r_{0})a_{z}-(ir_{1}+{}r_{2})a_{y}+h.c..
\end{split}
\end{equation}
And the quadrupolar spin spectral function $Q_{\mbox{\tiny{FQ3}}}(\bm{q},\omega)$ is the same as Eq.~\eqref{eq:ssffq3} and does not depend on the $\bm{d}$ vector.

$\bm{\mbox{AFQ3\ phase}}\quad$ The forms of $\bm{Q}$ operators for sublattice 1(2) are the same as Eq.~\eqref{eq:fq3q} (Eq.~\eqref{eq:fq1q}) with additional sublattice index. Thus the quadrupolar spin spectral function $Q_{\mbox{\tiny{AFQ3}}}(\bm{q},\omega)$ is the same as Eq.~\eqref{eq:ssfafq3}. and does not depend on the $\bm{d}$ vector.

\section{Connection with $t$-$J$-$V$ model.}
\label{sec:tjv}
Finally, we would like to point out that the $SU(2)\times U(1)$ model can be mapped to the $t$-$J$-$V$ model in one dimension, which is exactly solvable in the supersymmetric point \cite{sarkar91,sarkar95}.
We use $S_{\gamma}$ to denote the effective spin for $SU(2)_{\gamma}$ symmetry. 
Then the local spin state $|z\rangle$ is $SU(2)_{\gamma}$ invariant and belongs to $SU(2)_{\gamma}$ irreducible representation (IR) $S_{\gamma}=0$, while $|x\rangle$ and $|y\rangle$ belong to $SU(2)_{\gamma}$ IR $S_{\gamma}=\frac{1}{2}$. 
Therefore, $|z\rangle$ can be treat as a ``hole" state, and the $SU(2)\times U(1)$ symmetric model defined in Eq.~\eqref{eq:H123} can be mapped to the $t$-$J$-$V$ model, which reads\cite{SU3ULSL,SU3ULSS,tjv}
\begin{equation}\label{eq:tJV}
\begin{split}
H=&-t\sum_{j,a}{}P\psi^{\dagger}_{j,a}\psi_{j+1,a}P+h.c\\
&+\sum_{j}P\left(J{}\vec{s}_{j}\cdot\vec{s}_{j+1}+Vn_{j}n_{j+1}\right)P,
\end{split}
\end{equation}
where $P$ projects out states with double occupancy and electron spin $\vec{s}_{j}$ and $n_{j}$ on site $j$ are defined as
\begin{equation}
\vec{s}_{j}\equiv\sum_{a,b}\psi^{\dagger}_{j,a}\vec{\sigma}_{ab}\psi_{j,b},\ n_{j}\equiv\sum_{a}\psi^{\dagger}_{j,a}\psi_{j,a}.
\end{equation}
The spin-1/2 indices, $a,b=\uparrow,\downarrow$. 
Letting $|\uparrow\rangle$ and $|\downarrow\rangle$ correspond to $|x\rangle$ and $|y\rangle$, and the ``hole" state correspond to $|z\rangle$ in spin-1 system, we can establish a mapping form the $t$-$J$-$V$ model to the $SU(2)\times U(1)$ model defined in Eq.~\eqref{eq:H123} through
\begin{equation}
\begin{array}{l}
K_{1}=J/4,\  K_{2}=t/2,\  K_{3}=V/3.
\end{array}
\end{equation}
Hamiltonian given in Eq.~\eqref{eq:tJV} is equivalent to the $SU(3)$ symmetric model when $J=2t$ and $V=3t/2$. And the supersymmetric $t$-$J$-$V$ model can be realized when $K_1=K_2=-K_3/3$.


\bibliography{refer}

\end{document}